%% file: main.tex
\documentclass{article}

\usepackage{arxiv}

\usepackage[utf8]{inputenc} 
\usepackage[T1]{fontenc}    
\usepackage{hyperref}       
\usepackage{url}            
\usepackage{booktabs}       
\usepackage{amsfonts}       
\usepackage{nicefrac}       
\usepackage{microtype}      
\usepackage{lipsum}

\usepackage{acronym}
\usepackage{amssymb}
\usepackage{amsfonts}
\usepackage{array}
\usepackage{graphicx}
\usepackage{textcomp}
\usepackage{multirow}
\usepackage{chemformula}
\usepackage{siunitx}
\usepackage{soul}
\usepackage{enumitem}
\usepackage{rotating}
\usepackage{setspace}

\newcommand{\bparagraph}[1]{\vspace{1.05mm}\noindent\textbf{#1}}

\newcommand{\pmtuf}{\ch{PM_{0.1}}}
\newcommand{\pmt}{\ch{PM_{2.5}}}
\newcommand{\pmten}{\ch{PM_{10}}}

\input{acronyms.tex}

\title{Low-Cost Outdoor Air Quality Monitoring and Sensor Calibration: A Survey and Critical Analysis}

\author{
  Francesco Concas \\
  Department of Computer Science \\
  University of Helsinki \\
  P.O. 68 (Pietari Kalmin Katu 5) \\
  \texttt{francesco.concas@helsinki.fi} \\
   \And
  Julien Mineraud \\
  Department of Computer Science \\
  University of Helsinki \\
  P.O. 68 (Pietari Kalmin Katu 5) \\
  \texttt{julien.mineraud@helsinki.fi} \\
   \And
  Eemil Lagerspetz \\
  Department of Computer Science \\
  University of Helsinki \\
  P.O. 68 (Pietari Kalmin Katu 5) \\
  \texttt{eemil.lagerspetz@cs.helsinki.fi} \\
   \And
  Samu Varjonen \\
  Department of Computer Science \\
  University of Helsinki \\
  P.O. 68 (Pietari Kalmin Katu 5) \\
  \texttt{samu.varjonen@helsinki.fi} \\
   \And
  Xiaoli Liu \\
  Department of Computer Science \\
  University of Helsinki \\
  P.O. 68 (Pietari Kalmin Katu 5) \\
  \texttt{xiaoli.liu@helsinki.fi} \\
   \And
  Kai Puolamäki \\
  Institute for Atmospheric and\\ Earth System Research (INAR) \\
  Department of Computer Science \\
  University of Helsinki \\
  P.O. 68 (Pietari Kalmin Katu 5) \\
  \texttt{kai.puolamaki@helsinki.fi} \\
   \And
  Petteri Nurmi \\
  Department of Computer Science \\
  University of Helsinki \\
  P.O. 68 (Pietari Kalmin Katu 5) \\
  \texttt{ptnurmi@cs.helsinki.fi} \\
   \And
  Sasu Tarkoma \\
  Department of Computer Science \\
  University of Helsinki \\
  P.O. 68 (Pietari Kalmin Katu 5) \\
  \texttt{sasu.tarkoma@helsinki.fi} \\
}

\begin{document}
\maketitle

\begin{abstract}
\input{abstract.tex}
\end{abstract}


\input{sections/introduction.tex}
\input{sections/scope.tex}
\input{sections/pipeline.tex}
\input{sections/sensors.tex}
\input{sections/data.tex}
\input{sections/mlmodels.tex}
\input{sections/performance.tex}
\input{sections/roadmap.tex}
\input{sections/conclusion.tex}

\bibliographystyle{unsrt}  
\bibliography{references}

\end{document}

%% file: acronyms.tex
\acrodef{ANN}{Artificial neural network}
\acrodef{AAP}{Ambient air pollution}
\acrodef{AB}[AdaBoost]{adaptive boosting}
\acrodef{AQI}{air quality index}
\acrodef{AQS}{air quality sensor}
\acrodef{BAM}{beta attenuation monitors}
\acrodef{BFGS}{Broyden-Fletcher-Goldfarb-Shanno}
\acrodef{BMU}{best matching unit}
\acrodef{CO}{carbon monoxide}
\acrodef{CO2}[CO\textsubscript{2}]{carbon dioxide}
\acrodef{CDF}{cumulative distribution function}
\acrodef{CPC}{condensation particle counter}
\acrodef{CvMAE}{coefficient of variation of the mean absolute error}
\acrodef{DiSC}{Diffusion size classifier}
\acrodef{DL}{Deep learning}
\acrodef{DT}{Decision tree}
\acrodef{EPA}{United States Environmental Protection Agency}
\acrodef{EC}{electrochemical}
\acrodef{FAC2}{factor of 2 measure}
\acrodef{FFNN}{feedforward neural network}
\acrodef{GAM}{Generalised additive model}
\acrodef{GB}{gradient boosting}
\acrodef{GC}{Gas Chromatograph}
\acrodef{GD}{gradient descent}
\acrodef{GIS}{geographic information system}
\acrodef{GP}{Gaussian process}
\acrodef{GSN}{Global Sensor Network}
\acrodef{GMR}{geometric mean regression}
\acrodef{H2O}[H\textsubscript{2}O]{Water}
\acrodef{IID}{independent and identically distributed}
\acrodef{LR}{Linear regression}
\acrodef{LM}{Levenberg--Marquardt}
\acrodef{MLR}{multivariate linear regression}
\acrodef{LSP}{light-scattering particle sensors}
\acrodef{LSTM}{long-short term memory}
\acrodef{LUR}{land-use regression}
\acrodef{MAE}{mean absolute error}
\acrodef{MAPE}{mean absolute percentage error}
\acrodef{MBE}{mean bias error}
\acrodef{MEM}{Micro electromechanical}
\acrodef{MRE}{mean relative error}
\acrodef{MSE}{Mean square error}
\acrodef{ML}{machine learning}
\acrodef{MLP}{multilayer perceptron}
\acrodef{MOS}{metal oxide semiconductor}
\acrodef{M5P}{M5 model tree}
\acrodef{NARX}{nonlinear autoregressive exogenous model}
\acrodef{NDIR}{non-dispersive infrared}
\acrodef{NMHC}{non-methane hydrocarbon}
\acrodef{NO2}[NO\textsubscript{2}]{nitrogen dioxide}
\acrodef{NO}{nitrogen oxide}
\acrodef{NOx}[NO\textsubscript{x}]{a combination of \ac{NO} and \ac{NO2}}
\acrodef{O3}[O\textsubscript{3}]{ozone}
\acrodef{OPC}{optical particle counter}
\acrodef{PM25}[PM\textsubscript{2.5}]{PM\textsubscript{2.5}}
\acrodef{PCA}{Principal component analysis}
\acrodef{PID}{photo-ionisation detector}
\acrodef{PM}{particulate matter}
\acrodef{PM1.0}[PM\textsubscript{1.0}]{particulate matter 1.0}
\acrodef{PM2.5}[PM\textsubscript{2.5}]{particulate matter 2.5}
\acrodef{PM10}[PM\textsubscript{10}]{particulate matter 10}
\acrodef{POI}{point of interest}
\acrodef{RAMP}{Real-time Affordable Multi-Pollutant}
\acrodef{ReLU}{rectified linear unit}
\acrodef{RNN}{recurrent neural network}
\acrodef{RF}{random forest}
\acrodef{RH}{relative humidity}
\acrodef{RBF}{radial basis function}
\acrodef{RMSE}{root-mean-square error}
\acrodef{SGD}{stochastic gradient descent}
\acrodef{SMPS}{scanning mobility particle sizer}
\acrodef{SOM}{self-organising map}
\acrodef{SO2}[SO\textsubscript{2}]{sulphur dioxide}
\acrodef{SOC}[SoC]{System-On-a-Chip}
\acrodef{SVM}{support vector machine}
\acrodef{SVR}{support vector regression}
\acrodef{T}{temperature}
\acrodef{TEOM}{Tapered Element Oscillating Microbalance}
\acrodef{TDNN}{time delay neural network}
\acrodef{UFP}{ultrafine particle}
\acrodef{UV}{ultraviolet}
\acrodef{WHO}{World Health Organization}
\acrodef{XGB}[XGBoost]{extreme gradient boosting}

%% file: abstract.tex
The significance of air pollution and the problems associated with it are fueling deployments of air quality monitoring stations worldwide.
The most common approach for air quality monitoring is to rely on environmental monitoring stations, which unfortunately are very expensive both to acquire and to maintain. Hence environmental monitoring stations are typically sparsely deployed, resulting in limited spatial resolution for measurements.
Recently, low-cost air quality sensors have emerged as an alternative that can improve the granularity of monitoring.
The use of low-cost air quality sensors, however, presents several challenges: they suffer from cross-sensitivities between different ambient pollutants; they can be affected by external factors, such as traffic, weather changes, and human behavior; and their accuracy degrades over time.
Periodic \textit{re-calibration} can improve the accuracy of low-cost sensors, particularly with machine-learning-based calibration, which has shown great promise due to its capability to calibrate sensors in-field.
In this article, we survey the rapidly growing research landscape of low-cost sensor technologies for air quality monitoring and their calibration using machine learning techniques.
We also identify open research challenges and present directions for future research.

%% file: sections/introduction.tex
\section{Introduction}
\label{sec:introduction}

Air pollution is one of the most significant environmental challenges of our time.
According to the \ac{WHO}, in 2016, air pollution was linked to over 4.2 million deaths per year (11.6\% of all deaths), with mortality in low and middle-income countries particularly heavily affected by air pollution \cite{WorldHealthOrganization2016}.
Besides having a direct effect on mortality, air pollution is strongly associated with a broad spectrum of acute and chronic diseases, including cardiovascular diseases \cite{Brook2010, Goldberg2011, Andersen2012}, lung diseases \cite{Gehring2010, Andersen2011, Andersen2012a}, several types of cancer \cite{Goldberg2011, Raaschou-Nielsen2011, saber2012}, and even conditions affecting cognitive capabilities and the central nervous system \cite{Volk2013, Bos2014, zhang18cognitive}.
Air pollution is also a significant economic burden worldwide, with estimates suggesting that the world spends 2--5\% of overall GDP on treating diseases linked with air pollution~\cite{oecd2016economic}.
The severity of air pollution is exacerbated by ever-increasing urbanization, with estimates suggesting that 96\% of the world's population lives in areas where air pollution exceeds safe limits \cite{Lewis2016}.

Understanding the characteristics of pollutants in urban environments is essential for counteracting problems linked to poor air quality and for assessing the effectiveness of initiatives designed for tackling it.
This need for detailed air quality information is driving deployments of air quality monitoring technology worldwide, particularly in metropolitan regions.\footnote{\url{https://www.hindustantimes.com/delhi-news/delhi-gets-18-more-monitoring-stations-to-keep-tab-on-air-quality/story-kBtKpMeuPyz0KgOeDB1z9M.html}} \footnote{\url{http://www.baaqmd.gov/about-air-quality/air-quality-measurement/ambient-air-monitoring-network}}
\footnote{\url{http://www.chinadaily.com.cn/china/2016-02/22/content_23595631.htm}} Traditionally, air pollutant concentrations are monitored using professional air quality monitoring stations that meet strict accuracy criteria~\cite{Berkovicz1996, Vardoulakis2005, kulmala2018build}.
Such stations are highly accurate but also very expensive, with the cost of a single station often reaching hundreds of thousands or even million dollars~\cite{lagerspetz19feasibility}.
Operating such stations is also costly, requiring periodic maintenance from specially trained engineers.
Due to their high deployment and operating costs, professional stations are deployed sparsely, with most metropolitan regions only having a single measurement station.
While in line with official recommendations, such density is not sufficient, as even a single city block can witness significant variations in pollutant concentrations.
For example, congested traffic corridors, such as intersections or bus stops, tend to have significantly higher pollution concentrations than areas around them~\cite{apte2017high, moore2012air}.
To accurately assess the health and environmental risks of pollutants, it is also necessary to understand the chemical composition of pollutants, which varies depending on the season and characteristics of industry and traffic in the region~\cite{bell2009hospital, wang2003intercomparison, wang2005ion}.
For these reasons, accurate monitoring of air pollution inside metropolitan regions would require deploying hundreds or even thousands of air quality monitoring stations.
In contrast, the WHO recommends deploying one air quality monitoring station per square kilometre\footnote{\url{https://m.economictimes.com/small-biz/startups/newsbuzz/making-sense-of-air-quality-using-sensors/amp_articleshow/69262232.cms}}, whereas the EU clean air directive suggests one station per approximately\footnote{The recommended density varies across pollutants, with \num{200000} being the average density across all pollutants.} \num{200000} inhabitants~\cite{eu-2008-50}.

\textit{Low-cost air quality sensors} costing less than \$\num{10000} have recently emerged as a way to reach higher density deployments and achieve a higher spatial resolution in air quality monitoring~\cite{Spinelle2017, Szulczyski2017, Hasenfratz2012}.
Low-cost sensors are typically small in size, making it possible to deploy them densely as part of the urban infrastructure.
For example, low-cost sensors have been deployed onto light poles and public transport vehicles~\cite{ArrayOfThings18, Li2012}.
The main drawback of low-cost air quality sensors, however, is that their accuracy tends to be poor compared to professional monitoring stations~\cite{Bart2014, Cross2017, Masson2015a, Morawska2018}.
Indeed, measurements provided by low-cost sensors can vary significantly and have poor correspondence with professional-grade monitoring stations~\cite{Borrego2016}, with their performance best suited for specialized tasks where exact measurements are not required, such as detecting pollution hotspots~\cite{lagerspetz19feasibility}.

The accuracy of low-cost sensors can be improved through periodic re-calibration, with a single calibration cycle improving accuracy for up to a fortnight, before drift~\cite{Jiao2016} and other errors~\cite{Morawska2018} start to decrease accuracy.
Periodic calibration alone, however, is insufficient, since sensors are vulnerable to cross-sensitivities between different pollutants~\cite{Cross2017} and variations in atmospheric conditions, with temperature, humidity, and wind direction being examples of factors that influence sensor performance~\cite{Masson2015a}.
The calibration process is also highly time-consuming and laborious, which makes it unfeasible for large-scale deployments~\cite{Ramanathan2006}.

\textit{Machine-learning-based calibration} has recently emerged as a potential solution for improving the generality of calibration techniques and reducing work effort in the calibration process.
The general idea of these approaches is to co-locate low-cost sensors in proximity of a professional station, which is used as a reference, and to train a model that can correct the error of the low-cost sensors using weather measurements and other sources of information.
While several solutions for machine-learning-based calibration have been proposed~\cite{Maag2016, Zimmerman2018, DeVito2009}, the overall research landscape around machine-learning-based calibration is not sufficiently understood.
Indeed, there is limited information on which methods are best suited for tackling the research challenges posed by low-cost air quality monitoring stations, how to best evaluate low-cost sensor calibration techniques, and which are the other major research challenges in the area.

In this article, we contribute by surveying and critically analyzing the current research landscape on \textit{low-cost outdoor air quality monitoring stations} and their \textit{calibration using machine learning techniques}.
We focus specifically on low-cost technology aimed at {\em improving the resolution} of monitoring and {\em increasing the density} of deployments.
Previous surveys on air quality monitoring (see \autoref{ssec:surveys}) either focus on covering specific sensor technologies or on dealing with a specific measurement challenge without comprehensively reviewing the entire research landscape.
Besides reviewing existing work, we perform a rigorous analysis of the field to highlight significant open research challenges to establish a path forward.

%% file: sections/scope.tex
\section{Scope of the Survey}
\label{sec:scope}

Research on low-cost air quality sensing has been recently gaining momentum, as sensor technology has matured to a point where increasingly large-scale deployments are possible.
For example, Cheng et al.~\cite{cheng2019ict} consider a testbed consisting of \num{1000} low-cost sensors deployed in Beijing, and Motlagh et al.~\cite{motlagh2020towards} present a testbed with \num{100}+ sensors located in three different districts in Helsinki, Finland.
This gain in momentum also reflects in the number of published research papers on the calibration of low-cost sensors, with a query for \texttt{low-cost air quality calibration} returning over \num{180000} results on Google Scholar.
Despite this increase, the research challenges in the field are still poorly understood, and there is a lack of critical surveys assessing the state of the art.
Indeed, existing surveys mostly focus on individual pollutants and specific parts of the processing pipeline, without covering issues surrounding machine-learning-based calibration\footnote{In this article, we use \textit{calibration} exclusively to refer to the use of machine learning to improve the quality and accuracy of sensor measurements. The alternative to machine-learning-based calibration is metrological calibration, where the response of the device taking measurements is adjusted to match a reference signal.
Metrological calibration is typically conducted in carefully controlled laboratory conditions, whereas machine-learning-based calibration operates using measurements collected from a real-world deployment.} of low-cost air quality sensors in depth.
Our survey addresses this gap, providing a thorough review of the research field and performing a rigorous gap analysis to identify significant open research challenges in the field.

\subsection{Related Surveys}
\label{ssec:surveys}

\autoref{tab:existing-surveys} summarizes previous surveys having partially overlapping scopes with our work.
In terms of sensor technology, several existing surveys focus on specific types of sensors or technology.
However, these surveys have not addressed the suitability of different technologies for large-scale air quality monitoring or how technology affects the processing pipeline.

In terms of operations performed on sensor devices, only a small number of previous surveys exist.
These predominantly focus on a specific research challenge or specific parts of the pipeline.
For example, Morawska et al.~\cite{Morawska2018} provide an overview of deployments of low-cost air quality sensors but do not cover other parts of the processing pipeline.
Gama et al.~\cite{Gama2016} provide a general review of concept drift without focusing specifically on outdoor calibration.
As part of their studies on sensor calibration, Liu et al.~\cite{Liu2017a} and Zheng et al.~\cite{Zheng2018} survey studies on the calibration of optical \ac{PM} sensors.
Our survey supplements these surveys by providing an overview of approaches and research challenges in the calibration process of outdoor air quality sensors.
Closest to our work, Maag et al.~\cite{Maag2018survey} provide an overview of low-cost sensor calibration, focusing on different sources of error.
While the scope of our survey is similar, we address the field from a different angle, focusing on the operations needed to implement the calibration pipeline and the effect that different sensing technologies have on this pipeline.
Our survey also has a broader coverage of the calibration pipeline, thus supplementing the survey by Maag et al.~\cite{Maag2018survey}.
Having an in-depth understanding of the different sensor technologies and algorithms, as well as their advantages and disadvantages, is essential for developing best practices, identifying the most important research challenges, and advancing research in the field of low-cost air quality sensors outdoor calibration.

Finally, there have been surveys about application areas that require fusing air quality information from several sensors, such as how to use air quality information to generate spatiotemporal pollution maps~\cite{Hoek2008, Johnson2010}.
These surveys focus on what to do with air quality information produced by calibration models and thus supplement our survey, which addresses the operations needed to calibrate air quality sensors and to provide better quality data.

\input{tables/existing-surveys.tex}

\subsection{Selection of Articles}

We determined which articles to include (and exclude from) the survey through a three-stage process.
In the first stage, an iterative search strategy was used to determine potentially relevant articles to be included.
We first identified an initial set of articles using searches with a small set of keywords on Google Scholar, IEEE Xplore, and ACM Digital Library.
The following keywords were used: \textit{air quality}, \textit{sensors}, \textit{low cost}, \textit{machine learning}.
We complemented the results with follow-up works from most prominent researchers.
Once we formed the initial set, we searched for articles citing them or that were cited by the articles in this set.
We continued this process until we did not find more articles.
In line with the interdisciplinary nature of the topic, we carried out searches separately within computer science and atmospheric sciences publications.

In the second stage, during a preliminary screening phase, we pruned the articles found by the search engines by labeling them as potentially relevant or irrelevant by one of the researchers contributing to the survey.
We preserved any article relating to the scope of the survey.

In the third stage, once we identified the articles, we filtered them by asking one of the researchers that contributed to the survey to read the articles and to present the main findings for the other authors.
We then decided through a majority decision whether the article was within the scope of the survey.
In this survey, we aimed to focus on applying machine learning for the re-calibration of low-cost sensors.
Hence, we selected only papers that used machine learning in their re-calibration process.
However, we also evaluated several papers specific on low-cost sensors implementations to provide insights into the different sensing technologies in air pollution monitoring.

%% file: tables/existing-surveys.tex
\begin{table*}
    \centering
    \caption{Existing related surveys}
    \label{tab:existing-surveys}
    \resizebox{\linewidth}{!}{
    \begin{tabular}{l l}
        \toprule
         Scope & Survey\\\midrule
         \multirow{8}{*}{Sensors}
         & Gas sensor technologies~\cite{Wang2010,Li2012,Baron2017,Morawska2018}\\
         & MOS sensors~\cite{Ornek2012}\\
         & NDIR sensors~\cite{Dinh2016}\\
         & Portable sensors~\cite{Thompson2016,Spinelle2017a}\\
         & Wearable sensors~\cite{Maag2018}\\
         & Commercial sensors~\cite{Aleixandre2012,Szulczyski2017}\\
         & Low-cost sensors quality~\cite{Borrego2016,Borrego2018}\\
         & Usability of low-cost \acp{AQS} for atmospheric measurements~\cite{Alastair2018}\\
         \hline
         Deployment & Cities and projects~\cite{Morawska2018}\\
         \hline
         \multirow{2}{*}{Calibration}
         & Adaptation to drift~\cite{Gama2016}\\
         & Optical PM sensors~\cite{Liu2017a,Zheng2018}\\
         & Error sources in calibration~\cite{Maag2018survey} \\
         \hline
         \multirow{3}{*}{Integration}
         & Air quality sensor networks~\cite{yi2015survey}\\
         & Land-use regression models~\cite{Hoek2008,Johnson2010}\\
         & Satellite-based estimations~\cite{Streets2013,Duncan2014}\\
         \bottomrule
    \end{tabular}
    }
\end{table*}

%% file: sections/pipeline.tex
\section{Low-Cost Air Quality Sensing Pipeline}
\label{ref:pipeline}

\begin{figure}
    \centering
    \includegraphics[width=.8\linewidth]{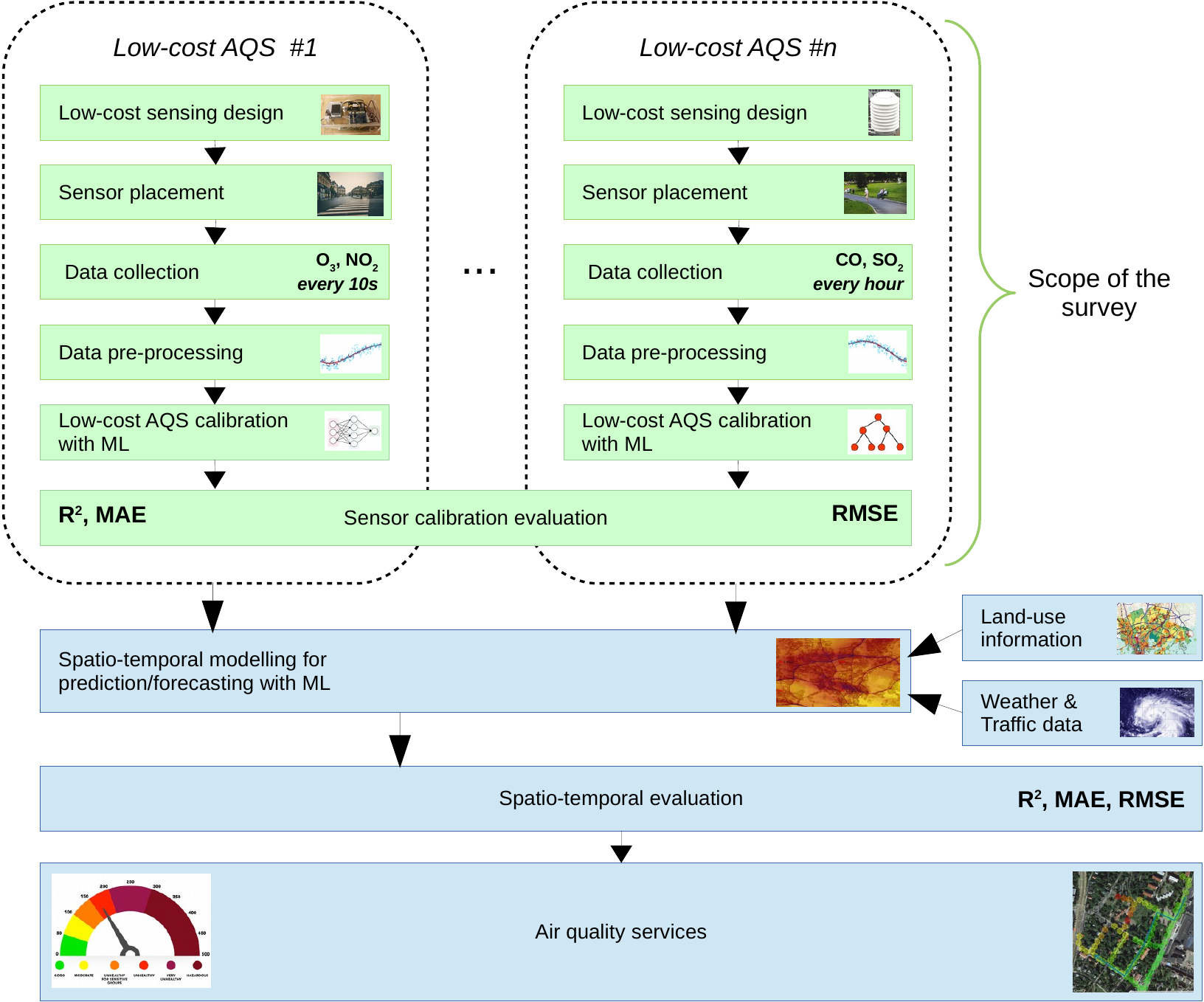}
    \caption{A reference calibration pipeline for low-cost air quality sensors. Different services may use variants of the pipeline.}
    \label{fig:pipeline}
\end{figure}

Low-cost air quality sensing follows a typical machine learning pipeline for sensor data, illustrated in \autoref{fig:pipeline}.
Within the pipeline, we can separate two types of operations: \textit{per-device} operations, which need to be performed separately for each sensor, and \textit{integration} operations, which combine data from multiple sensing units.
Note that this distinction does not mean that the devices cannot collaborate, but instead, whether a specific operation needs to be applied to each device or the aggregated data.
Indeed, calibration functions usually use measurements from multiple different sensors to construct the underlying calibration model, but, at runtime, the model is applied separately to each device, making this a per-device operation.
The specifics of the pipeline can vary for different services, even if the general structure remains similar. For example, local pollution prediction requires a different time scale and update interval than daily city-level pollution estimates \cite{donnelly2015, liu2017urbanair, Cheng2014}.
In this survey, our focus is on per-device operations, as several existing surveys are focused on application domains that are related to integration operations~\cite{yi2015survey, Hoek2008, Johnson2010, Streets2013, Duncan2014}.
Below we briefly give an overview of the different steps in the sensing pipeline.

\subsection{Per-Device Operations}

The low-cost air-quality sensing pipeline includes six operations that address issues related to individual devices: 1) the design of low-cost sensing units, 2) deciding where the devices should be deployed, 3) the actual data collection, 4) pre-processing, 5) \ac{ML}-based calibration and 6) the evaluation of that calibration.
Applications, such as prediction and real-time maps, can then be built on the results of per-device data.

\bparagraph{Low-cost Sensing Design.} A sensing unit can be seen as a low-cost monitoring station integrating one to several sensors, each measuring a specific pollutant or environmental variable.
In current air quality sensing research, the design of the sensing units is largely overlooked.
Indeed, most of the research relies on off-the-shelf sensing units and focuses on other parts of the pipeline.
However, the choice of sensor models and the overall design of the sensing unit can have a significant impact on how these sensors produce data. 
For example, as we discuss in \autoref{sec:sensors}, some sensors need to be heated, while others are sensitive to temperature changes.
When sensors for several pollutants are integrated into the same sensing unit, heating some sensors can result in significant inaccuracies by sensors sensitive to temperature changes.
Another concern is related to variations in sampling rates that can make it difficult to synchronize measurements from different sensors or to relate the measurements with real-world events.
We discuss the properties of low-cost gas and particle matter sensors in \autoref{sec:sensors}.

\bparagraph{Sensor Placement.} The sensors can be mobile, mounted onto vehicles, carried as personal sensors (i.e., wearables), or deployed to specific areas of cities.
The benefit of mobile sensors is that their measurements can cover a large area.
However, this can result in high sparsity, as many areas will only have a small set of measurements.
Another challenge with mobile sensors is how to manage them and how to ensure they are operational.
Instead of relying on mobile sensors, most real-world deployments rely on sensors placed in fixed positions for an extended time.
Examples include the Barcelona Lighting Masterplan\footnote{\url{http://ajuntament.barcelona.cat/ecologiaurbana/en}} and the Chicago Array of Things\footnote{\url{https://arrayofthings.github.io}}.
In current research, the deployment of sensors is largely driven by practical constraints, such as availability of power and availability of locations where the airflow to the sensor remains unobstructed.
We briefly discuss sensor placement in \autoref{sec:sensors} and refer to the survey by Morawska et al.~\cite{Morawska2018} for a more thorough overview of existing sensor deployments.

\bparagraph{Data Collection.} This step consists in collecting air pollution data, i.e., the concentration of selected air pollutants and environmental parameters.
Pollutants require different collection densities, therefore, not every pollutant needs to be collected by all devices~\cite{targino2016hotspots}.
For example, temperature and humidity have similar patterns within one region, whereas pollutants resulting from vehicles can fluctuate significantly, even within a small region.
Data collection is discussed in \autoref{sec:calibration}.

\bparagraph{Data Pre-processing.} Examples of pre-processing operations include synchronization of different measurements and removal of erroneous measurements from periods where the device is compromised.
For instance, the devices could have been operating in extreme conditions which are not supported by the sensors (e.g., high temperatures), the measurement units may be clogged, or power spikes can disrupt the functionality of a sensor.
Other common pre-processing techniques include interpolation to achieve a consistent sampling rate and aggregation to reduce sampling rate, e.g., when low-cost sensors produce data at a higher rate than professional-grade reference stations.
State-of-the-art low-cost air quality research typically considers measured pollutant concentrations and environmental variables (such as temperature and relative humidity) as features.
Pre-processing is discussed in detail in \autoref{sec:calibration}.

\bparagraph{Calibration Based on Machine Learning.} This step consists in the application of machine learning techniques to calibrate the measurements of the low-cost sensors, and it is the main focus of our survey.
We critically compare existing machine learning solutions for the calibration of low-cost air quality sensors and identify the main advantages and disadvantages of the methods.
Machine-learning-based calibration is discussed in detail in \autoref{sec:calibration}.

\bparagraph{Sensor Calibration Evaluation.} The final step related to a single sensing unit is the performance evaluation of their calibration.
We discuss this step in detail in \autoref{sec:performance} with an emphasis on the selection of performance measures and test data length.

\bparagraph{Integration Operations.} Air quality data from a single sensor is limited in usefulness as it captures pollutant concentrations only within a small region.
In practice, to create high-quality spatiotemporal air quality information with fine granularity, we need to combine data from numerous sensing units.
This operation is conducted in the seventh step, which includes spatiotemporal modeling and fusion with additional sources of information, including but not limited to land-use, weather, and traffic data.
The eighth step includes the performance evaluation of the models produced by the previous step.
Finally, the ninth and last step is the production of air quality services built upon high-quality air pollution models.
Such services include more advanced \ac{AQI}~\cite{Olstrup2019} models and green path routing~\cite{Hasenfratz2015} to enhance the quality of life of citizens.
We note that while such services are not strictly part of the calibration pipeline, air quality services can have contrasting accuracy, resolution, and other requirements.
These requirements, in turn, affect the calibration pipeline, and thus the services are intrinsically linked with it.

%% file: sections/sensors.tex
\section{Low-cost air quality sensors}
\label{sec:sensors}

Low-cost solutions for air pollution monitoring typically consist of sensing units that package together multiple low-cost air quality sensing components.
Besides the components responsible for monitoring pollutants, sensing units can incorporate other components, such as power sources, processing units, local data storage, networking interfaces, and atmospheric sensors (e.g., temperature, relative humidity, and wind direction)~\cite{Dutta2009}.
Individual components usually cost between \$\num{20} and \$\num{100}.
However, a complete low-cost sensing unit typically costs upwards of \$\num{500}~\cite{Castell2017}.
In this article, we use the term \textit{sensor} to refer to a single sensing component responsible for measuring a (single) pollutant or environmental variable, whereas we use the term \textit{sensing unit} to refer to the overall device used to collect measurements for the individual studies in question.

In this section, we survey the most commonly used sensor technologies employed by low-cost air quality sensing units and discuss how their characteristics affect in-field calibration.
Characteristics of components and the overall design of a sensing unit can have a significant impact on resulting measurements, and hence need to be carefully chosen to avoid negatively influencing other parts of the air quality monitoring pipeline.
Indeed, as we discuss in Sec.~\ref{ssec:comparing}, the best model for a pollutant rarely is the same that works well for another pollutant.
For this reason, it is important to understand the characteristics of the underlying sensor technology, as well as the key advantages and disadvantages of the sensor technology.
Before delving into the details of the sensor technologies, we briefly discuss the two different types of pollutants considered in air quality research.

\subsection{Types of Pollutants}
\label{ssec:pollutants}

Air quality sensing research typically categorizes pollutants as gaseous or particulate matter, according to the composition of the pollutant~\cite{Rinne12pollution}.
Commonly considered gaseous pollutants include carbon monoxide (\ch{CO}), carbon dioxide (\ch{CO2}), ozone (\ch{O3}), nitrogen oxides (\ch{NO_{x}}) and sulfur oxides (\ch{SO_{x}}).
Carbon dioxide monitoring is restricted to indoor environments.
In outdoor monitoring, the most common gases are the ones belonging to prominent air quality indexes, such as EPA in the USA or the Air Quality Index of China.
Specifically, these indexes include the following gases: sulfur dioxide (\ch{SO2}), ozone (\ch{O3}), carbon monoxide (\ch{CO}) and nitrogen dioxide (\ch{NO2}).
Particulates (or aerosols), on the other hand, refer to tiny particles of solid or liquid compounds suspended in gases.
The source of particulates can be natural (e.g., dust or sea spray) or caused by human activity (e.g., burning of fossil fuels or wood, dust from roads and tires, and power plants).

Gas sensors can be typically tailored to support different gases by changing (parts of) the sensing materials or operating parameters of the sensing unit.
Particulate matter sensors, in contrast, only monitor the extent of particulates in the air without being able to identify their exact source or composition.
However, particulate matter sensors can be categorized based on the size of the particles they can monitor, with fine (\pmt{}) and coarse-grained (\pmten{}) being the most common categories in low-cost air quality research and belonging to all major air quality indexes.
With some sensor technologies, it is also possible to detect so-called ultra-fine particulates (\pmtuf{}).
However, these mostly require expensive professional-grade instruments and thus are rarely considered in low-cost sensing research.

\subsection{Sensing Gaseous Pollutants}

Within low-cost sensing units for gaseous pollutants we can identify four main types of sensing technologies: \ac{MOS} sensors, \ac{EC} sensors, \ac{NDIR} sensors, and \ac{PID} sensors.
In this section, we describe the key properties of these technologies.
We discuss \ac{MOS} and \ac{EC} sensors in more detail as they are the least expensive and most widely used technologies in low-cost sensing units.
NDIR and PID sensors have higher costs than \ac{MOS} and EC sensors (\$\num{500}--\$\num{5000}); hence, they are rarely used in low-cost sensing units.
We note that there are also other technologies for monitoring concentrations of gaseous air pollutants, but these have even higher costs, making them unsuitable for low-cost sensing units.
For example, gas chromatography (GC) sensors cost between \$\num{15000} and \$\num{100000}.

\subsection*{Solid-State Metal Oxide Sensors}

\ac{MOS} sensors are a popular sensor technology for monitoring gas concentrations of several pollutants, such as \acp{NMHC}, \ac{CO}, \ac{CO2}, \ac{NO2}, \ac{NOx} and \ac{O3}~\cite{DeVito2008, DeVito2009, Hasenfratz2012a, Piedrahita2014, Oletic2015, Saukh2015, Spinelle2015, Spinelle2017, Maag2018}.
\ac{MOS} sensors consist of a heating element and a semiconducting metal oxide sensing element.
The heater warms the surface of the sensing element up to 300--500\textdegree C, which is then able to detect gases through a chemical reaction occurring on its surface.
This reaction causes a change in the electrical conductivity of the sensing element, which can be monitored using an external circuit to measure the detected gas level~\cite{Ornek2012}.

\begin{description}
    \item[Advantages:] \ac{MOS} sensors are very low-cost and compact.
    Furthermore, these sensors have a high sensitivity and can even reach sub-parts per billion (ppb) sensitivity for some gases.
    \ac{MOS} sensors also have a short response time, i.e., they can produce data at a high sampling frequency.
    Other advantages of \ac{MOS} sensors include their long lifespan and resilience against extreme weather conditions.
    Indeed, \ac{MOS} sensors can operate in high temperature and humidity environments, making them well-suited for (very) long-term deployments.
    
    \item[Disadvantages:] \ac{MOS} sensors have several disadvantages that can affect other parts of the processing pipeline.
    First, while they are resilient, their sensitivity is affected by atmospheric conditions and other gases.
    In terms of calibration, this implies that inputs from temperature and relative humidity sensors are needed while calibrating \ac{MOS} sensors.
    Another disadvantage of \ac{MOS} sensors is that they tend to have low accuracy and are subject to drift~ \cite{Wang2010}, which requires them to be re-calibrated often to maintain good quality data outputs~\cite{Tsujita2005, Masson2015}.
    Drift manifests as a reduced measurement accuracy, which results from a decrease in the conductance of the sensing element over time~\cite{romain2010long}.
    The response of MOS sensors also depends on humidity, with higher humidity resulting in higher sensor error~\cite{sohn2008characterisation}.
    Finally, \ac{MOS} sensors require access to a sufficiently large power source due to their need to power an electric heater.
\end{description}

\ac{MOS} sensors are well-suited for low-cost sensing units due to their low cost and high resilience against environmental conditions.
However, their high power requirement and sensitivity to environmental conditions are a concern.
High power requirements make \ac{MOS} sensors better suited for deployments where a fixed power supply is available than for deployments requiring battery power.
As an example of the use of \ac{MOS} sensors for air quality monitoring, Hasenfratz et al.~\cite{Hasenfratz2012} used a \texttt{MiCS-OZ-47} sensor as part of a participatory air quality monitoring campaign, where they connected the sensor to a separate battery pack and used a smartphone to store and transmit measurements.
The authors evaluated the impact of the sensor on the battery, which resulted in an estimated operation time of 50 hours when using a separate battery.
Burgues et al.~\cite{burgues2018} state that shutting down the heating elements for some time and then heating them in cycles rather than continuously can reduce the power consumption of gas sensors up to 90\%.
However, the energy savings come at the cost of reduced measurement accuracy, suggesting that the duty cycle of the sensors needs to be taken into consideration when building calibration models.
In particular, two MOS sensors with similar specifications but different duty cycles are likely to have different accuracy characteristics that need to be taken into account when training and sharing measurements or calibration models across devices.
Another alternative is to design the duty cycle to maximize measurement quality.
For example, if multiple measurements are collected in the same cycle and the first few measurements are dropped, the resulting measurements are more accurate than when only a single measurement is collected in each cycle.~\cite{burgues2018}.

Recently, there have been advances in the synthesis of \ac{MOS} gas sensing materials that can ensure sensitivity remains high even when air humidity increases~\cite{vasiliev2018}.
These materials are not yet available in mass quantities, hence they are not used by current low-cost sensors.

\subsection*{Electrochemical Sensors} 

\ac{EC} sensors are the other popular low-cost technology for monitoring gas concentrations.
EC sensors have been used to monitor \ac{CO}, \ac{NO}, \ac{NO2}, \ac{O3} and \ac{SO2} \cite{Nikzad2012, Saukh2015, Oletic2015, Spinelle2015, Spinelle2017, Esposito2016, Hu2016, Zimmerman2018}.
These sensors detect gases by oxidation-reduction reactions, employing electrodes separated by an electrolyte substance, such as mineral acid.
The working electrode is in contact with both the electrolyte and the ambient air, which is monitored via a porous membrane.
The reaction produces an electrical current between the electrodes, which can be measured from the outer pins of the sensor.

\begin{description}
    \item[Advantages:] Similarly to \ac{MOS} sensors, \ac{EC} sensors are inexpensive, have high sensitivity (in parts per billion (ppb) levels for some gases), and good specificity.
    Compared to \ac{MOS} sensors, the main benefit of EC sensors is their lower power draw as they do not require powering an electric heater.
    Another advantage of EC sensors is that their sensitivity is less affected by temperature and humidity than with \ac{MOS} sensors.
    
    \item[Disadvantages:] The main drawback of EC sensors is that the speed of the chemical reaction depends on the operating conditions.
The operating range of the sensors is also dependent on the characteristics of the chemicals used in the sensor and tends to be narrower than in MOS sensors.
EC sensors also have a shorter lifespan than MOS sensors, with the overall duration depending on the amount of pollution they are exposed to.
    As an example, the Alphasense \ch{NO2} sensor used by Castell et al.~\cite{Castell2017} has a lifetime of $2$--$5$ years, whereas MOS sensors can last $10$ years or even longer.
    For example, Romain and Nicolas~\cite{romain2010long} used MOS sensors to collect data over a $7$ year period.
    EC sensors are also less resilient to weather conditions than \ac{MOS} sensors.
    A combination of low humidity and high temperature is particularly problematic to EC sensors, as it can dry out the sensor's electrolyte and break the sensor.
    Another drawback with EC sensors is that other gases may interfere with the measurements, even though they are less sensitive than \ac{MOS} sensors~\cite{Thompson2016}.
\end{description}

\ac{EC} sensors are well-suited for low-cost deployments, as the sensors are inexpensive and their performance is less affected by temperature and humidity variations than \ac{MOS} sensors.
The accuracy of \ac{EC} sensors tends to be good, as long as the weather conditions fall within their operational range.
Examples of research relying on EC sensors include Nikzad et al.~\cite{Nikzad2012} and Wei et al.~\cite{Wei2018}.
Ultimately the choice between MOS and EC sensors depends on the goals of the deployment.
In areas with fairly stable temperature and weather conditions, EC sensors are well suited due to their better specificity and accuracy.
Also, the scale and intended duration of the deployment affect the choice of technology, with MOS sensors being capable of supporting deployments lasting $10$ years or more, whereas EC sensors tend to be limited to deployments lasting few years at most.
In terms of calibration, the two technologies are reasonably similar, with the main difference being how often the calibration functions need to be re-trained and validated.

\subsection*{Non-dispersive Infrared}

\ac{NDIR} sensors consist of an infrared light source, an atmospheric sampling chamber, an optical filter, and a detector.
When a gas passes through the chamber, the light emitted by the infrared source travels through it, and some frequencies get absorbed depending on the gas.
The rest of the light hits the optical filter and the detector, which outputs the frequencies through an electrical current.
\ac{NDIR} sensors have been mostly used to detect \ac{CO2} concentrations \cite{Piedrahita2014, Spinelle2015, Spinelle2017, Zimmerman2018}.
However, they can be used to detect also other gases through changes in the wavelength of the light.

\begin{description}
    \item[Advantages:] They are simple and require little power, and small units also are available.
    They have a long lifespan (not degraded by exposure to gases) and require only little maintenance.
    
    \item[Disadvantages:] They have high detection limits, i.e., cannot measure small pollutant concentrations, and they are susceptible to spectral interference from different gases as well as water~\cite{Thompson2016}.
    \ac{NDIR} sensors are also subject to drift \cite{Dinh2016} and they cost considerably more than \ac{MOS} or EC sensors (up to a \num{10}-fold increase in price).
\end{description}

Even considering the disadvantages, the long lifespan of \ac{NDIR} sensors makes them a good choice for long-term deployments in dry areas.
The high cost and high detection limit mean that they are better suited for sparser deployments than what most low-cost sensing aims at accomplishing.

\subsection*{Photo-ionisation Detectors}

\ac{PID} sensors operate by illuminating compounds using high energy UV photons.
The process results in compounds becoming ionized as they absorb the UV photons and results in an electrical current that can be captured by a detector inside the sensor.
The greater the concentration of the measured component in the compound, the more ions are produced, and the greater the current.

\begin{description}
    \item[Advantages:] \acp{PID} are very sensitive and have a short response time.
    They are small in size and weight, and they have low energy requirements~\cite{Aleixandre2012, agbroko2017}.
    \item[Disadvantages:] \acp{PID} affect all molecules whose ionization potential is lower than the UV light affecting them, which means \acp{PID} are not specific to a particular pollutant~\cite{Aleixandre2012}.
    \acp{PID} are sensitive to high humidity levels or water vapor.
    They are not very low-cost (between \$\num{500} and \$\num{5000}), even if affordable compared to high-end monitoring stations~\cite{Spinelle2017a}.
    \acp{PID} are also subject to drift and need to be re-calibrated often (once per month).
\end{description}

The ability of \ac{PID} sensors to analyze samples of low concentration in ambient temperatures and pressures makes them suitable for deployment in a wide range of environments, which has made \ac{PID} sensors particularly well suited for the analysis of small particles and gases in controlled small-scale experiments.
However, the sensitivity to water and the relatively high cost make \ac{PID} sensors poorly suited for dense long-term deployments.

\input{tables/gas-sensing-technologies.tex}

Spinelle et al.~\cite{Spinelle2015, Spinelle2017} studied the performance of \ac{MOS} and \ac{EC} sensors for detecting \ch{O3}, \ch{NO}, \ch{NO2} and \ch{CO}.
The authors concluded that no significant differences in sensor outputs can be observed between the two sensor technologies.
In total, they evaluated 15 sensor models for the five gases.
However, their study included only five months of data, which is within the range of the lifespan of \ac{EC} sensors, therefore \ac{EC} sensor degradation effects were not considered.

Sensor mobility has been discussed in numerous studies that we review in this survey \cite{Cheng2014, Gao2016, Hasenfratz2012, Liu2017, Maag2016, Maag2018}.
For example, Hasenfratz et al.~\cite{Hasenfratz2012} simulated mobility by carrying experiments in a room with constant \ch{O3} concentration and artificial wind, generated using a table fan.
The authors found that, when \ch{O3} concentration is low, the wind does not affect the measurements much.
However, when \ch{O3} concentration is high, wind effects produce a measurement offset.
Therefore, they recommend shielding the sensor from the wind when moving relatively fast, for example, when riding a bicycle.

Generally, the choice between sensor technologies depends on the characteristics of the deployment.
MOS and EC sensors are cheapest and thus best suited for large-scale deployments, whereas PID and NDIR sensors are better suited to sparser deployments.
In terms of accuracy, MOS and EC sensors have comparable performance, with EC sensors being more energy-efficient, but MOS sensors being more resilient.
In particular, EC sensors do not require an electric heater, but they can break in high temperatures and low humidity conditions.
MOS sensors also are more durable than EC sensors.
In terms of calibration, both sensors are somewhat sensitive to weather conditions and concentrations of other gases, suggesting that these variables need to be incorporated into the calibration model.
In terms of sensor design, MOS sensors can cause cross-interference to other sensors by heating the air inside the sensing unit.
Thus, the sensing unit and the sensor cycles need to be carefully designed to minimize the risk of such effects.

\subsection{Particulate Matter Sensors}

Particulate matter sensors monitor particle density by assessing changes in the properties of air passing through the sensor unit.
Unlike gas sensors that can be tailored to detect different pollutants, particulate matter sensors cannot identify the exact composition of pollutants.
However, they can be adapted to recognize particles of different sizes.

Existing low-cost particulate matter sensing technology is predominantly based on \acp{DiSC} or \acp{LSP}.
\acp{DiSC} operate by charging air passing through the sensor and estimating particle density from the total electricity charge after applying different filtering operations on the charged air.
\acp{LSP} operate by using light scattering to estimate the density of the particles.
Traditional laboratory-grade instruments for particulate matter sensing are based on similar principles, but use additional components to improve detection accuracy.
For example, \acp{OPC} are high-quality variants of \acp{LSP} whereas \acp{CPC} use alcohol or water vapor to change the physical properties of particulates before passing them through an \ac{LSP} sensor~\cite{SChmoll2010, Sousan2016, Shao2017, Chen2016b}.
However, these sensors are typically bulky and more expensive than basic \acp{DiSC} and \acp{LSP}, and hence rarely used in low-cost air quality sensing.

At the top end of the scale, \ac{TEOM} sensors use changes in the oscillation frequency of a vibrating glass tube, and \ac{BAM} sensors use absorption of beta radiation for estimating particle density.
TEOM and BAM sensors are quite expensive, costing over \$\num{20000}~\cite{Shao2017}.
Another high-end option is scanning mobility particle sizer (SMPS), which estimates the size and concentration of particles using electrical mobility sizing to monodisperse the output, which is then monitored with a \ac{CPC} sensor.
In the following subsections, we describe \ac{DiSC} and \acp{LSP} in more detail since these technologies are the most affordable and most widely used in low-cost air quality sensing research.

\subsection*{Diffusion Size Classifiers}

\acp{DiSC} detect particles by applying electrical signals to identify physical changes on the sensing surface.
They contain an air inlet, a corona charger, an induction stage, a diffusion stage, and a backup filter.
When air enters the sensor, particles go through a diffusion charger that produces ions using a corona wire.
A small fraction of these ions attach to particles in the air, and an ion trap placed between the diffusion charger and the induction stage captures the remaining ions.
The particles then pass through the induction stage, where they produce a small electrical current that is proportional to their concentration.
After leaving the induction stage, the particles arrive at the diffusion stage that precipitates the particles to produce a small electrical current, which is proportional to their concentration.
Since the particles also have an induction effect in the diffusion stage, the current measured in the induction stage is subtracted from the current measured in the diffusion stage to compensate for the induction effect.
Larger particles not precipitated by the diffusion stage eventually reach the backup filter, which produces an electrical current proportional to their concentration~\cite{Fierz2011, Meier2012}.
This way \ac{DiSC} sensors can separate between different particle sizes.

\begin{description}
    \item[Advantages:] The sensitivity of \ac{DiSC} sensors is extremely high, and they have low power consumption.
Manufacturing costs are low when the sensors are manufactured in large bulks.
    \item[Disadvantages:] Manufacturing costs of \ac{DiSC} have a high upfront setup cost, as they require clean rooms and other special facilities.
    Therefore, the production and assembly of low quantities of sensors have a high unit cost ($\approx$ \$\num{10000}).
    Also, testing equipment for assessing the quality and performance of \ac{DiSC} sensors can be expensive.
    Another problem with \ac{DiSC} sensors is that the sensing area can become unclean, making it necessary to clean it frequently.
    Research on how to automatize the cleaning process, e.g., using oscillation and electrostatics, is actively pursued~\cite{Kim2018, Meier2012}.
\end{description}

\subsection*{Light-Scattering Particle Sensors}

\acp{LSP} are small, low-cost sensing units widely used to detect particulate matter~\cite{Cheng2014, Liu2017, Liu2018, Chowdhury2018, Kuula2017}.
They are composed of an air inlet, a light sensor, and a light source, usually infrared or laser.
When air enters the sensor, the light source is focused on a sensing point.
An infrared LED is positioned at a forward angle relative to a photodiode.
Particles passing through the light beam scatter light, which generates a measurable signal in the sensor circuitry.
The scattered light is focused onto the photodiode by a lens.
The sensors may contain a light scattering focusing lens and a focusing lens also for the infrared light source.
The resolution at which different particle sizes can be detected depends on the configuration of these lenses.
Finally, the sensor produces a signal that can be measured to estimate the number of particles in the air~\cite{Chowdhury2007}.

\begin{description}
    \item[Advantages:] \acp{LSP} are small and their cost is very low.
    The sensors mostly require very low power~\cite{Chowdhury2018}.
    \item[Disadvantages:] Light-scattering based instruments fail to detect very small particles~\cite{Fierz2011}.
    The sensor readings are also impacted by temperature and relative humidity, which means both temperature and humidity need to be measured.
    More expensive \acp{LSP} can use multi-angular light scattering to reduce the impact of environmental variables~\cite{Shao2017}.
\end{description}

\input{tables/pm-sensing-technologies.tex}

Table~\ref{tab:pm-sensing-technologies} summarizes the advantages and disadvantages of the two particulate matter sensing technologies used in low-cost air quality monitoring research.
Similarly to gas sensors, the optimal choice of sensing technology depends on the context of the deployment.
\ac{DiSC} sensors have high sensitivity and are mostly unaffected by weather or other environmental conditions.
However, they suffer from the need for regular maintenance to ensure the detection surface remains sufficiently clean.
\acp{LSP}, on the other hand, are susceptible to weather changes but require less maintenance, making them better suited for longer-term deployments.
While other technologies for particulate matter sensing exist, they are not well suited for low-cost deployments due to the higher cost of sensing equipment and larger size of the sensing units.

Weather affects the concentration of particles in the air, and thus the sensor measurements.
Therefore, particulate matter sensors generally require information about weather conditions regardless of whether the underlying sensor technology is sensitive to weather or not.
High winds can disperse particulate concentration, whereas high humidity causes particulates to cluster together, increasing its concentration.
The effect of temperature, however, is less well understood.
Zheng et al.~\cite{Zheng2013} found higher temperatures to lower particulate concentrations when humidity is low, and lower temperatures to decrease particle concentrations when humidity is high.
Wang et al.~\cite{Wang2015} studied the effects of temperature and humidity on low-cost PM sensors and found relative humidity to affect the accuracy of sensor technology.
For example, as water in the air absorbs infrared radiation, humidity can result in \acp{LSP} overestimating particle concentrations as light intensity is reduced.
The temperature was not found to directly impact the sensor technology, even if it affects the concentration of particles.

Besides weather, the concentration of particulate matter is affected by the extent of human activity within the area being monitored.
The higher the traffic density and the lower the fuel efficiency of the vehicles, the higher the concentration of particulates will be.
Note that the density is not solely a result of fuel burning as also tire and road wear produce particulates.
In terms of calibration, this implies that the context of the deployment needs to be taken into consideration as locations close to intersections are likely to have an increased particulate matter concentration than other nearby areas~\cite{Zheng2013}.
Several research initiatives have explored the possibility of mounting low-cost sensors on vehicles~\cite{Cheng2014, Gao2016, Hasenfratz2014}.
When the vehicles are moving, the input air flux is constantly changing, which can affect sensor accuracy.
In particular, the more air enters into the sensor, the more pollutants will be detected, even if the concentration of particulates would remain constant.
Besides vehicles, wind speed can trigger a similar effect.
To ensure that an accurate calibration model can be constructed, a possible solution is to include a GPS---or another sensor to estimate the velocity of the sensor unit at the time of measurement---and to use this information to compensate for the density of pollutants detected by the sensor~\cite{Gao2016}.

%% file: tables/gas-sensing-technologies.tex
\begin{table*}
    \centering
    \caption{Advantages and disadvantages of gas sensor technologies with respect to their use within low-cost sensor arrays}
    \label{tab:gas-sensing-technologies}
    \resizebox{\linewidth}{!}{
    \begin{tabular}{l c c c c c c c c c m{16em}}
        \toprule
        Type & Cost & Size & Lifespan & Sensitivity & Drift & Accuracy & Energy & Calibration & Response time & Known issues\\
        \midrule
        MOS & Low & Small & Long & High & Yes & Low & High & Frequent & Fast & Cross-sensitivity to humidity and other gases. Sensitivity reduced in high temperature.\\
        \hline
        EC & Low & Small & Short & High & \shortstack{2\%--15\% \\ per year}  & Good & Low & Reasonable & $\approx 120s$& Sensitive to temperature. Low humidity and high temperatures can cause the electrolyte of the sensor to dry out.\\
        \hline
        NDIR & High & Small & Long & High & (0.4 ± 0.4)\% & High & Low & Frequent 
        & $\approx 20s$ 
        & Spectral interference and high detection limit. Cross-sensitivity at least to water vapour. \\
        \hline
        PID & High & Small & Long 
        & High & \shortstack{20\% \\ in weeks }
         & High & low & Frequent & \shortstack{Fast \\ $\approx 1s$} & High sensitivity to high humidity levels or water vapour. \\
        \bottomrule
    \end{tabular}
    }
\end{table*}

%% file: tables/pm-sensing-technologies.tex
\begin{table*}[t]
    \centering
    \caption{Advantages and disadvantages of \ac{PM} sensor technologies with respect to their use within low-cost sensor arrays}
    \label{tab:pm-sensing-technologies}
    \resizebox{\linewidth}{!}{
    \begin{tabular}{l c c c c c c c >{\raggedright\arraybackslash}m{5em} c >{\raggedright\arraybackslash}m{15em}}
        \toprule
        Type & Cost & Size & Lifespan & Sensitivity & Drift & Accuracy & Energy & Calibration & Response & \multicolumn{1}{c}{Known issues} \\
        & & & & & & & & & time & \\
        \midrule
        \ac{DiSC} & High & Small & Good
        & Good & Yes & Good & High & Yearly (Factory),\linebreak Hourly (Software) & 1s & The instrument can produce wrong results if the incoming aerosol is highly positively charged. Cannot distinguish between narrow and broad particle size distributions \cite{Fierz2011}. \\
        \hline
        \ac{LSP} & Ultra-low & Small & Good & Poor & None & Low & Low & Frequent & $\approx 30s$  & Mixes all particle sizes, variation of air influx reduces accuracy. \\
        \bottomrule
    \end{tabular}
    }
\end{table*}

%% file: sections/data.tex
\section{Data Collection and Pre-Processing}
\label{sec:airquality}

Low-cost outdoor air quality monitoring commonly relies on sensors deployed as part of the urban infrastructure~\cite{Morawska2018}.
The most common approaches are to deploy the sensors as part of fixed infrastructure, such as street lights, or to mount the sensors onto vehicles, such as trams~\cite{Li2012}, garbage trucks~\cite{shirai2016toward}, or even Google Street View vehicles~\cite{apte2017high}.
Next, we briefly describe the characteristics of the data collected by low-cost sensor deployments and typical preprocessing operations performed on the measurements.

\subsection{Measurements}

Low-cost air quality sensors measurements can be interpreted as time-series data consisting of values of different pollutants and environmental variables.
The integration of several pollutants is necessary to capture the effects of cross-sensitivities~\cite{Cross2017}, whereas environmental variables are critical for accounting for differences in error as environmental conditions change.
As discussed in Section~\ref{ssec:pollutants}, the most common pollutants to consider are gaseous pollutants and particulate matter included in prominent air quality indexes.
In terms of environmental variables, temperature and relative humidity are the most common variables that need to be considered.
Wind speed is another variable that can influence pollutant concentrations.
However, wind speed is often difficult to measure with low-cost sensors, as it requires an unobstructed air intake, whereas other environmental variables can be more accurately and reliably measured with low-cost sensors.

\bparagraph{Reference Measurements.} Calibrating low-cost sensors with machine learning requires access to high-quality reference measurements to consider as ground truth.
The most common choice is to deploy low-cost sensors near a high-quality atmospheric station, and use the measurement of the station as ground truth~\cite{DeVito2009, Esposito2016, Spinelle2015, Spinelle2017}.
Another possibility is to use a high-quality mobile measurement laboratory deployed near the low-cost sensors~\cite{Zimmerman2018}.
Generally, the closer the low-cost sensor is to the reference station, the better an appropriate calibration relationship can be established.
In cases in which the ground truth is needed in multiple different locations, for example, when calibrating mobile sensors, other approaches must be used~\cite{motlagh2020towards}.
For example, public high-quality pollution data from official authorities can be used~\cite{Hasenfratz2012}.

\bparagraph{Temporal resolution.} The temporal resolution of measurements also influences the calibration process.
The resolution is governed by the sampling frequency of the sensors, which varies across different sensor technologies.
For example, as MOS sensors require heating, they have a slower sampling rate than sensors that can operate continuously---unless the heating element is run continuously, which would result in prohibitively high power drain.
In most of the studies surveyed for this article, the sampling rate is between 5 seconds and 20 seconds~\cite{Maag2016, Saukh2015, Zimmerman2018, DeVito2009}.
However, there are also studies with sampling intervals as high as one hour \cite{Borrego2018} or as low as ten milliseconds~\cite{Spinelle2015, Spinelle2017}.

\subsection{Characteristics of Air Quality Measurements}

Low-cost sensor measurements have some characteristics that need to be accounted for while designing calibration solutions.
These characteristics all affect the statistical properties of the measurements and can result in natural variations in measurements or from different sources of errors.
As they affect the measurements used as input for calibration algorithms, they need to be accounted for when selecting appropriate algorithms.
Below we briefly discuss the most important characteristics.

\bparagraph{Autocorrelation.} Air pollutant concentrations are known to have a strong spatiotemporal correlation with the weather.
Furthermore, seasonal patterns also have a significant influence on them~\cite{chock1975time,merz1972aerometric,salcedo1999time},
which means that the used calibration techniques need to be capable of dealing with autocorrelation.
The data used to test the generality of the model should also be sufficiently long-term to ensure the results are not overfitting to short-term correlations.

\bparagraph{Cross-sensitivities.} Measurements provided by low-cost sensors suffer from cross-sensitivities between pollutants~\cite{Cross2017}.
Measurements can also be affected by temperature, humidity, and wind direction~\cite{Masson2015a}.
In terms of calibration models, this means that the used techniques cannot assume the variables to be independent, but instead, they need to consider the values of environmental variables, and potentially also the values of other pollutants, as input.

\bparagraph{Drift.} Low-cost air pollutant measurements are vulnerable to \textit{drift} whereby the relationship between environmental variables and pollutants varies over time~\cite{Gama2016}.
For example, an analysis of metal oxide sensors (MOS; see Section~\ref{sec:sensors}) has demonstrated that the measurements can differ by over $200\%$ over time~\cite{romain2010long}.
The most common reason for drift is wear.
For example, metal oxide sensors are vulnerable to oxidation, which alters the conductivity of the sensing element and results in drift~\cite{barsan2007metal}, whereas light-scattering particle sensors are vulnerable to deposits forming on the lens of the optical sensor~\cite{austin2015laboratory}.
The traditional way to handle drift is to perform maintenance on the sensors, which requires cleaning and replacing components and re-calibrating the sensors in laboratory conditions.
For large-scale deployments, this is not feasible, and an alternative approach is re-training the calibration function to account for changes in the properties of the sensors.
The frequency of re-calibration depends on the characteristics of the deployment.
For example, light scattering particle sensor re-calibration frequency depends on the extent of pollutants.
The higher the amount of pollutants, the more often re-calibration is required.

\bparagraph{Concept Drift.} Besides being vulnerable to mechanical or chemical drift affecting the sensor hardware, field deployments are vulnerable to \textit{concept drift} where the statistical characteristics related to estimation of the target variables change over time.
These changes can be a result of persistent effects, such as clean air policies or changes in human consumption patterns~\cite{mcdonald2018volatile}, or temporal effects, such as forest fires, volcano eruptions, or other weather phenomena~\cite{radke1978airborne}.
In terms of calibration function, concept drift can be accounted for by adapting the underlying calibration function or re-training it, depending on the magnitude of changes.
De Vito et al.~\cite{devito2020robustness} discuss how the risk of concept drift can be recognized by analyzing the statistical difference in distributions of air pollution measurements.

\bparagraph{Height differences.} The most common placement of low-cost sensors is near the ground level without isolating them from the urban infrastructure.
Professional-grade measurement stations, on the other hand, are typically at least partially isolated from the urban infrastructure, and they have different air intakes located in different parts of the sensor.
For example, in Finland, reference stations are either in dedicated containers that can be near the ground or as part of separate measurement towers that are jointly responsible for weather and pollution measurements~\cite{kulmala2018build}.
Pollutant concentrations can vary significantly also with elevation.
For example, seasonality influences the elevation of the atmospheric mixing layer~\cite{tang2016mixing}, which in turn affects the extent of pollutants that can be captured~\cite{wagner2017influence}.

\subsection{Preprocessing}

The measurements from low-cost sensors typically require preprocessing before they can be used to capture a calibration model.
Below we briefly describe the most common operations.

\bparagraph{Resampling.} Before training a calibration model, measurements need to be resampled to a suitable temporal resolution.
A higher resolution implies more samples and a longer model training time, but it can improve the robustness of the model.
A common choice is to use a one-minute resolution~\cite{Spinelle2015, Spinelle2017, Esposito2016, Gao2016}.
Some studies use a much coarser resolution, such as one hour~\cite{DeVito2009, Borrego2018}, which is the standard resolution for deriving air quality index values~\cite{USEPA2016}.
Saukh et al.~\cite{Saukh2015} use a resolution in tens of seconds, while Hasenfratz et al.~\cite{Hasenfratz2012} and Maag et al.~\cite{Maag2016} use five seconds.

\bparagraph{Synchronization.} To minimize cross-sensitivities, sensor sampling intervals need to be interleaved.
For example, as MOS sensors require heating, they can influence measurements for temperature or other pollutants unless the heating period is sufficiently distant from the sampling period of other sensors.
To account for differences in sampling times, the measurements need to be synchronized during the modeling phase.
This can be accomplished using aggregation, e.g., using the mean value over a given data window, or interpolating values of different sensor units to have consistent timestamps.
In most cases, using a simple linear interpolation is sufficient, especially if the synchronization window is short.

\bparagraph{Smoothing and Filtering.} Pollutant measurements occasionally contain outliers that can significantly decrease the performance of the calibration model, unless the outliers are accounted for.
There are several reasons for outliers.
For example, sudden wind gusts can result in abnormal measurements, or the air intake of the sensing unit may get temporally obstructed.
Common ways to mitigate these issues are to use \textit{smoothing} or \textit{filtering}.
In the former case, the measurements are fitted to a model that ensures temporal consistency, whereas in the latter case, values appearing as abnormal or otherwise erroneous are removed.
As an example of the former, Cheng et al.~\cite{Cheng2014} smooth the sensor data through a signal reconstruction technique based on a bi-criterion problem with a quadratic smoothing function.
As an example of the latter, Hasenfratz et al.~\cite{Hasenfratz2014} used a three-phased filtering process.
First, each sensor computes its null-offset and uses it to calibrate the offset of the measured particle concentration.
Second, the sensing units check that the sensors are operating correctly and measurements from periods where sensor failures are detected are removed.
Third, measurements with poor location accuracy are removed from consideration.
Another example of filtering is proposed by Gao et al.~\cite{Gao2016} who also use GPS data to filter measurements.
Smoothing and filtering are popular techniques for preprocessing, and as such, are likely to be used in most studies.
However, most studies we surveyed for this article do not indicate which kind of preprocessing has been applied~\cite{Spinelle2015, Spinelle2017, Cordero2018, Zimmerman2018}.

%% file: sections/mlmodels.tex
\section{Machine Learning Calibration of Low-Cost Air Quality Sensors}
\label{sec:mlmodels}

\input{tables/cal_summary.tex}

Low-cost sensors increasingly rely on machine-learning-based calibration pipelines to improve the accuracy of sensor measurements.
Previous surveys have shown that domain characteristics need to be taken into account when building such calibration pipelines~\cite{Liu2017a, Zheng2018, Maag2018survey}.
In the following, we survey existing calibration pipelines for low-cost air quality sensor data and discuss how they address requirements stemming from the specifics of air quality measurements.
Generally, these approaches learn the relationship between a specific pollutant, as given by the low-cost sensor, and a ground truth value obtained from a reference station.
We first discuss issues that affect the choice of machine learning algorithms, after which we survey machine learning algorithms used in previous studies.
The different models and how they have been applied are summarized in Table~\ref{tab:cal_summary}.

\begin{figure}
    \centering
    \includegraphics[width=.55\linewidth]{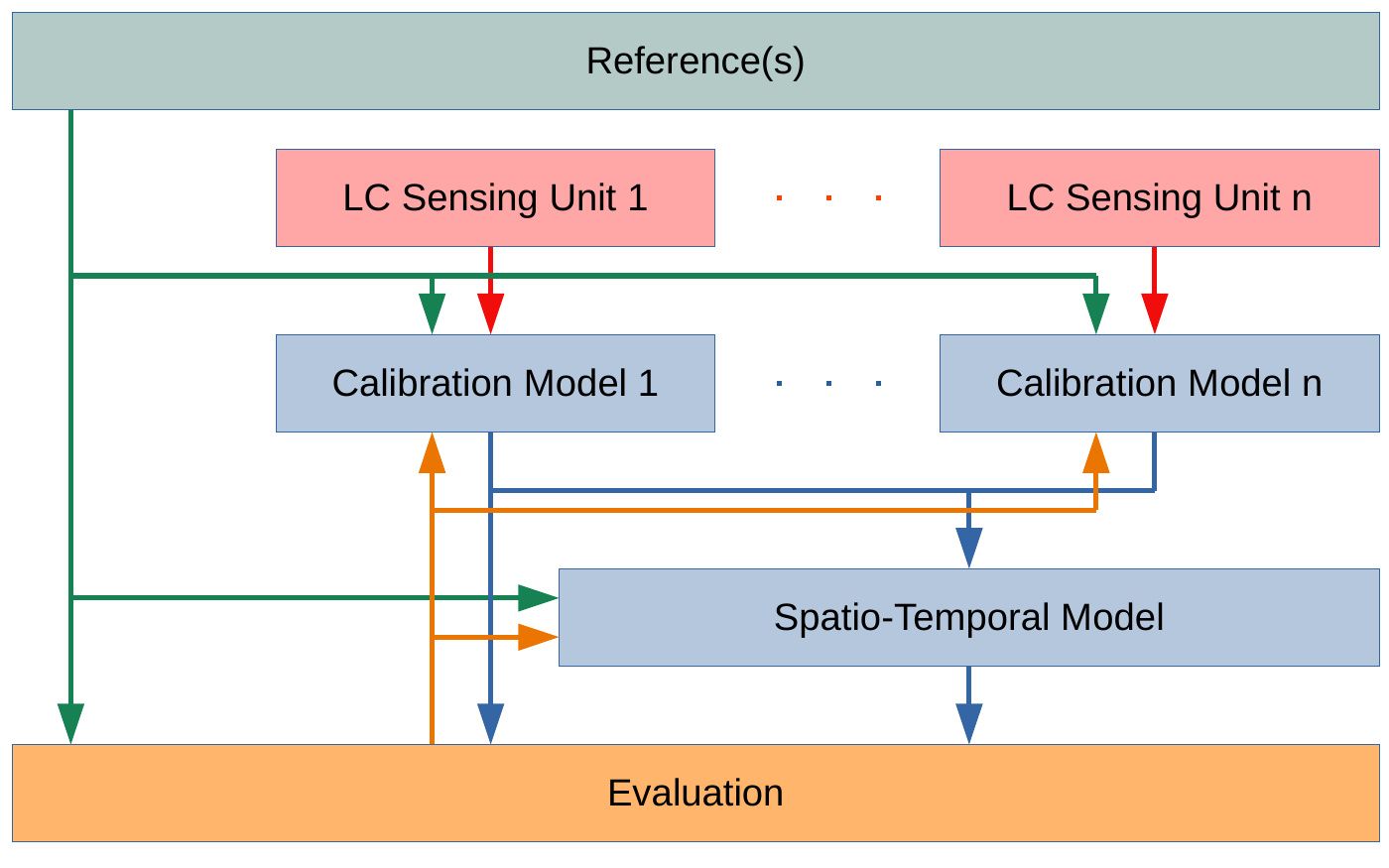}
    \caption{Data flow diagram of machine-learning-based calibration.}
    \label{fig:cal_loop}
\end{figure}

Figure~\ref{fig:cal_loop} presents a high-level overview of the (continuous) machine learning calibration process.
A reference station provides reference data to train calibration models that can correct the measurements of a particular low-cost sensing unit.
Note that each sensor (type) can have a different model, as shown in the figure.
The performance and usefulness of the models are evaluated against new measurements, similarly collected from low-cost sensor units and a reference station.
Once the error of a calibration model is sufficiently small, the corrected measurements can be used by many air quality sensing applications, such as pollution monitoring and prediction, and high definition pollution maps based on spatiotemporal models~\cite{Johnson2010, Eeftens2012, Beelen2013, Wang2013}.

\subsection{Issues for Calibration}
\label{sec:calibration}

\bparagraph{Generality.} Calibration models should be capable of operating under different conditions, including different geographic locations, across different seasons, and potentially also across variations in sensing units.
In practice, simultaneously achieving all these goals is not feasible.
A model might need to be periodically retrained to adjust for variations.
Different models might be needed for different seasons, hardware, or even locations.
In the simplest case, such models correspond to reparametrizations of the same type of model, whereas in some situations it may be necessary to use different classes of models altogether, e.g., a neural network in one setting and a linear regression model in another.

\bparagraph{Distribution of Air Quality Data.} Many common machine learning techniques, such as common regression models, have been designed for data that is independent and identically distributed (i.i.d.), or at least close to being.
As discussed in the previous section, this rarely is the case for air quality data, as measurements contain significant temporal correlations, and the values of different pollutants and environmental variables are dependent on each other.
Therefore, calibration models that avoid making strong assumptions about the distribution of measurements are likely to perform and generalize better.

\bparagraph{Interpretability.} The most powerful machine learning models often are black boxes that hide their internal logic from the user~\cite{Lipton2018, Guidotti2018}.
While a black-box model may produce more accurate estimates, its nature makes it difficult or impossible for the user to understand how the model works or why it produces a particular estimate.
If the calibrated measurements will be used to make decisions that have, e.g., economic, legal, or safety consequences, then black-box approaches may be unacceptable.
Another concern with black-box models is that errors in the calibration models they capture may be difficult to rectify.
Indeed, a complex learning algorithm may, e.g., use unexpected correlations and features in the training data to obtain the estimates, which may lead to unexpected and counter-intuitive calibration behavior in real-life applications.
For a thorough discussion on black-box models and issues with them, we refer to Guidotti et al.~\cite{Guidotti2018}.
The alternative to black boxes is to use white boxes, models of which the performance can be easily understood and explained.
In practice, however, the number of parameters influences the interpretability of models, and most common white-box models become gray boxes that can be only partially explained.

\bparagraph{Optimization criteria.} Most common machine learning algorithms operate by minimizing an objective function that specifies a \textit{loss} between the output of the machine learning model and the desired output.
In the case of calibration, the loss function measures differences between the low-cost sensor and the reference station.
Generally, the cost function needs to be chosen so that it represents the goal of the application.
For example, if we wish to know whether the measured concentration is of the correct order of magnitude, mean squared error would be a good choice as a cost function, as it penalizes large errors more than small ones.
On the other hand, if we wish to optimize the median performance of the algorithm, we could use mean absolute error instead.
In some cases, we may be interested in other kinds of objective functions.
For example, for detecting drastic short bursts in the concentration of a pollutant, the objective function can be mapped into classification error, where the different classes represent the severity of the pollutant.

\bparagraph{Amount of Data.} Choosing the right machine learning model for a calibration problem depends on the available amount of air quality measurements.
The amount of air quality measurements, in turn, is linked with the generality and complexity of the model.
If the model is simple, it requires relatively few data points for training, but it may not fit the data well.
This phenomenon is known as \textit{underfitting}.
Conversely, if the model is complex, it will approximate the function to predict the data better, but it will also need a larger training dataset to avoid \textit{overfitting}, i.e., fitting the training dataset well but having poor performance on unseen data.
Generally, the more complex the model, the longer the training dataset should be.
Training data should also be sufficiently long to ensure that the model has enough information to learn how the cross-sensitivities between the sensors affect their response.
A longer period of training data generally implies a longer training time, which results in less data being available for validating and testing the model.

The choice of test data is critical for ensuring the usefulness of the calibration model.
Standard machine learning evaluation techniques, such as selecting a subset of all data as test data, or using cross-validation, are not suitable for air quality calibration due to the nature of the measurements.
Indeed, these evaluation techniques can result in significant correlations between training and testing data, which would result in the calibration model overfitting on the temporal structure of the measurements.
Optimal evaluation of air quality calibration models is currently an open issue as the data should cover a sufficiently long period, different pollutant concentrations, and different environmental conditions.
From the studies surveyed for this article, it is difficult to estimate average lengths of training or testing periods since many studies do not report them exactly.
In the studies that report the length, the length of training data spans from 2.1 days to 8 months, whereas the length of the test data spans from 36 hours to 1.25 years.

\bparagraph{Computational Time and Complexity.} Recent developments with networking (e.g., 5G and edge computing), artificial intelligence, and sensor technology are paving the way to ever-increasing scales for deployments, even for massive-scale deployments that integrate thousands or tens of thousands of sensors~\cite{motlagh2020towards}.
Unlike in traditional sensor network deployments, which rely on tens or hundreds of low-cost sensors and sparse reference station data, in massive-scale deployments, performing the calibration at remote infrastructure becomes difficult, especially if real-time (or near real-time) information is required.
For such deployments, the calibration model needs to be sufficiently lightweight in terms of computational time so that a pre-trained model can be used to adjust the measurements directly on the sensor device. This is particularly important for battery-powered monitoring stations, such as those carried by citizens, where any saving from the calibration and network costs offsets the energy cost of the air quality sensors themselves. Note that, even in this case, some of the measurements need to be transmitted to a remote server, edge deployment, or the cloud to support re-training the calibration model.
Even if the scale of the deployment is smaller, time requirements of machine learning techniques also have some influence on the choice of machine learning models for calibration, as they influence the resolution of measurements and potentially the energy drain of the sensing units running the calibration models. For fixed deployments where the low-cost sensors have mains power access, computational time and complexity are less of an issue as there is no need to adjust network or CPU power and calibration can be performed in a remote location.

Most machine learning models are slow to train and fast to use while predicting the values of new measurements.
The time requirements of the prediction phase should be sufficiently low so that the calibration model can be used to correct any new measurements from the low-cost sensing unit.
In practice, any machine learning model can achieve this since most low-cost air quality sensors require sufficient air intake before they can take measurements.
Indeed, the temporal resolution of measurements is usually in minutes (or once per minute) rather than in seconds.
For prediction, a more critical concern is memory and storage complexity.
Traditional machine learning techniques, such as linear models, random forests, and support vector regression, have a reasonably small model size, but emerging techniques, such as deep learning, often have large a model size, incorporating tens or even hundreds of thousands of parameters.
Storing and loading such models on the sensing units may become a bottleneck, especially on sensing units that have been designed to operate for a long time.
For model training, time complexity influences how often re-calibration can be performed, as well as the overall system architecture.
Simpler models, such as linear models or support vector regression, can be efficiently trained even on low-cost sensing units.
On the other hand, more complex models, such as artificial neural networks or deep models, can be too computationally heavy for the sensing units.
When training cannot be performed on the sensing units, sufficient computing and networking infrastructure need to be available.

\subsection{Linear Models}

\ac{LR} is the simplest machine learning regression model, and it is based on the linear equation.
In the case of a single (input) feature, this model is usually referred to as univariate \ac{LR} or just \ac{LR}.
In the case of more than one feature, this model is usually referred to as \ac{MLR}.
\ac{LR} models have widely been used as calibration methods for air quality monitoring~\cite{Hasenfratz2012, Lin2015, Saukh2015}, or as a baseline for comparing the calibration performances of more complex approaches~\cite{Spinelle2015, Spinelle2017, Zimmerman2018}.
\ac{LR} has also been used as a pre-calibration method, with the output of the model fed to other methods~\cite{Cordero2018}.
As discussed earlier, low-cost sensors are vulnerable to cross-sensitivities and meteorological conditions, which renders simple \ac{LR} models insufficient for most environments~\cite{Spinelle2015, Spinelle2017, Zimmerman2018}.
\ac{MLR} has shown improvement in performance~\cite{Spinelle2015, Spinelle2017, Saukh2015, Maag2016, Cordero2018}, because the model can learn cross-sensitivities between different pollutants and meteorological conditions.

When the relationship between the features and the target pollutant is not strictly linear, data transformation can be used to improve the accuracy of \ac{LR} models.
For instance, Liu et al. \cite{Liu2017} argued that \ac{LR} using a logarithmic function reacts better to the cross-sensitivity between PM\textsubscript{2.5} and wind interference.
A variation of \ac{MLR} called \ac{GMR} has been tested by Saukh et al.~\cite{Saukh2015}.
Their comparison with a conventional \ac{MLR} model shows that \ac{GMR} is less vulnerable to concept drift than conventional \ac{MLR} models.

Conventional linear models assume input data to be independent and identically distributed.
As this assumption rarely holds with air pollutants, linear models can only reach modest calibration performance.
However, a major benefit of linear models is that they are easily interpretable since the model corresponds to a (hyper)plane fitted on data.
Hence the parameters have an intuitive geometric interpretation and linear models can be considered as white-box models.
Due to their simplicity, linear models require less training data than more complex models. However, to ensure good performance, the training data needs to be sufficiently representative of the actual distribution of pollutants.
For example, the temporal range needs to be sufficiently long to capture potential effects resulting from seasonal variation, differing weather, or other factors.
Furthermore, linear models are very quick, both to train and to predict with.
This holds also for air quality data as the optimal weights that minimize the cost function can be directly computed, instead of being approximated through iterations, as the number of features is small.

\begin{description}
    \item[Advantages:] Simple to define, and trivial to find the optimum weights.
    \item[Disadvantages:] The low-cost \ac{AQS} calibration problem is too complex for this model~\cite{Spinelle2015, Spinelle2017, Zimmerman2018}.
    \ac{LR} cannot automatically find all the cross-sensitivities between the various pollutants, and in some cases, the function that models some relations is not linear.
    In such cases, the features need to be manually transformed to allow \ac{LR} to fit them.
\end{description}

\subsection{Support Vector Regression}

\ac{SVR} is one of the most popular techniques for modeling \textit{non-linear} relationships between input features and the output variable (i.e., calibrated air pollutant).
The general idea in \ac{SVR} is to find the hyperplane that represents the minimum distance between itself and the data points.
\ac{SVR} is well suited for non-linear data, as it uses a so-called kernel function to map non-linear data to a higher-dimensional space, in which the model can then fit a linear hyperplane.
In air quality calibration, \ac{SVR} has mainly been used for comparison against other techniques.
For example, Cordero et al.~\cite{Cordero2018} compare \ac{SVR}, random forest, and \acp{ANN} for calibrating \ch{NO2} measurements.
The authors report mixed results, with \acp{SVR} outperforming the other models in some cases but having worse performance in other cases.

\ac{SVR} models are more complex than linear models, hence, they require more training data.
In terms of runtime, they are fast to compute, even if less efficient than linear models.
\acp{SVR}, like linear models, assumes that the data is independent and identically distributed.
The interpretability of \ac{SVR} models depends on the kernel used, with a linear kernel resulting in a white-box model, but non-linear kernels effectively turning the model into a black-box that is difficult to interpret.

\begin{description}
    \item[Advantages:] Model training is defined as a convex optimization problem, for which there are efficient solutions to find the optimal parameters.
    \ac{SVR} uses a kernel to transform the input data, enabling capturing nonlinear relationships between input features and calibrated pollutants.
    \ac{SVR} incorporates a regularization parameter, which makes it possible to control under and overfitting.
    There are many efficient, mature, and easy-to-use \ac{SVR} implementations.
    \item[Disadvantages:] Sensitive to kernel parameters and assumes data to be independently and identically distributed.
Poor interpretability for nonlinear kernels.
\end{description}

\subsection{Decision Trees and Random Forests}

\acp{DT} are another common off-the-shelf machine learning technique.
In a \ac{DT}, every node of the tree has a conditional check, and each branch corresponds to an outcome of the check.
While determining the value of a new measurement, we progress through the tree, starting from the root node and following the branches, until we reach a leaf node.
The value of the leaf node is then used as the outcome of the calibration.
Each check in the \ac{DT} corresponds to a rule that can be used to subdivide the measurements into an increasingly smaller range of values.
Thus, unlike linear models and \acp{SVR}, \acp{DT} do not fit a model or function to the data, hence they can support both linear and nonlinear relationships.
To the best of our knowledge, \acp{DT} have found limited use in air quality calibration.
They have only been used for calibrating \ch{CO}, and even then only as a baseline for other methods~\cite{Hu2017}.

When applied to complex problems, \acp{DT} can become highly complex and overfit on the training data.
These issues can be mitigated using \acp{RF} which combine multiple simple \acp{DT} into a single powerful model.
\ac{RF} is an example of the \textit{bagging} technique, which operates by creating different subsets of the training data, learning separate models for each subset, and aggregating the outputs of all these models together while predicting unseen data.
The main disadvantage of \ac{RF} models is that they can be difficult to interpret.
Indeed, while the outputs of individual \acp{DT} can be easily interpreted, understanding the joint effect of tens or even hundreds of \acp{DT} is much more complex.
Similarly, the number of \acp{DT} to integrate into the model influences overall runtime and performance.
Having a small number of \acp{DT} as part of the RF model is efficient to learn; however, it can result in the model underfitting the data, whereas integrating a large number of trees increases the model size and training data requirement.
\acp{RF} models have been widely applied to the calibration of sensors for many pollutants, such as \ch{CO}~\cite{Zimmerman2018,Borrego2018}, \ch{CO2}~\cite{Zimmerman2018}, \ch{NO2} and \ch{O3}~\cite{Zimmerman2018,Cordero2018,Borrego2018}, \ch{NO}, \ch{SO2}, PM\textsubscript{2.5} and PM\textsubscript{10}~\cite{Borrego2018}.
Lin et al.~\cite{lin2018calibration} propose a hybrid model that combines \ac{LR} and \ac{RF} to simultaneously learn linear and non-linear relationships.

\begin{description}
    \item[Advantages:] Reduces overfitting by training different models on different artificial datasets generated from the original dataset.
    The training of different models can be parallelized.
    \acp{DT} and \acp{RF} do not require choosing the function for non-linear problems, which is an advantage in our scenario.
    \item[Disadvantages:] \acp{DT}, when used for regression, can have a very high depth unless properly regularized.
    The number of models, and therefore of generated datasets adds additional complexity to training the parameters correctly.
    The computational needs of the \acp{RF} are higher than a single \ac{DT}.
    \acp{RF} can be considered gray-box or even black-box algorithms, depending on the number of models that are integrated.
\end{description}

\subsection{Boosting Algorithms}

Similarly to bagging, \textit{boosting} algorithms operate by creating subsets of the training data and learning a different model on each subset.
The difference to bagging is that boosting trains all the weak models sequentially, aggregating them into a single strong model, instead of running each model separately and aggregating their outputs.
Boosting, when training a new weak model, also takes into account the success of the previously trained weak model, and weights the training data accordingly.
Example of boosting algorithms are \ac{AB}~\cite{Freund1996}, \ac{GB}~\cite{Friedman2001}, and \ac{XGB}~\cite{Chen2016a}.
In the context of air quality monitoring, boosting algorithms have been used for predicting \pmt{}~\cite{Johnson2018}.

\begin{description}
    \item[Advantages:] Reduces overfitting by training different models on different artificial datasets generated from the original dataset.
    No need to choose the function for non-linear relationships.
    When training the new weak learners to combine into the strong learner, the model takes into account the success of the previous weak learner and weighs data accordingly.
    \item[Disadvantages:] Interpretability and complexity depend on the number of learners that are aggregated together, with a higher number resulting in better performance, but higher complexity and reduced interpretability.
    Risk of overfitting when the number of learners to aggregate grows large.
\end{description}

\subsection{Artificial Neural Networks}

\acp{ANN} are a popular machine learning technique for modeling time series data.
ANNs consist of a set of nodes called \textit{neurons} grouped into \textit{layers}; their structure has been initially motivated by the structure of the human brain~\cite{Jain1996}.
The first layer of the model is called the input layer, and the last is referred to as the output layer.
Intermediate layers between the input and output layers are referred to as hidden layers.
Depending on the number of intermediate layers, the network is called either \textit{shallow} or \textit{deep}.
For the calibration of low-cost air quality models, the most common approach is to use a shallow \ac{FFNN}, with most models containing one or two hidden layers.
This type of \acp{ANN} have been applied to the calibration of sensors for measuring \acp{NMHC} \cite{DeVito2008}, PM2.5 \cite{Cheng2014, Gao2016}, CO \cite{DeVito2009}, CO\textsubscript{2} \cite{Maag2018}, NO \cite{DeVito2009, Esposito2016, Cordero2018}, NO\textsubscript{x} \cite{Esposito2016}, NO\textsubscript{2} \cite{DeVito2009, Esposito2016, Cordero2018}, and O\textsubscript{3} \cite{Maag2018, Cordero2018}.
Spinelle et al.
\cite{Spinelle2015, Spinelle2017} use an \ac{ANN} model based on \acp{FFNN} applied to the calibration of CO, CO\textsubscript{2}, NO, NO\textsubscript{2} and O\textsubscript{3}. Lee et al. \cite{Lee2020} design a combination calibration process selecting the prevailing calibration models between LR and ANN based on those two models' distribution of residuals.

In an \ac{ANN}, each neuron contains a function, named \textit{activation function}, which is applied to the data that it receives as input, to produce an output value.
Common activation functions include the \ac{ReLU} function, the sigmoid function, and the hyperbolic tangent function.
Similarly to \ac{LR}, the activation functions consist of some weights that need to be set so that the final output of the model is as close as possible to the ground truth.

Just a few of the studies that we survey discuss the used activation function of the neurons.
Some report using the hyperbolic tangent function, or a variation of it~\cite{DeVito2008, DeVito2009}.
Spinelle et al.
\cite{Spinelle2015, Spinelle2017} use different activation functions in different layers in their \ac{ANN} model.
They also report trying the \ac{RBF} function but discarding it because it did not yield good results.

In terms of performance, there have been studies that compare ANNs against regression models~\cite{Spinelle2015, Spinelle2017, Maag2018, Cordero2018}.
The conclusions from these studies are mixed.
Some studies report a better performance from \acp{ANN}~\cite{Spinelle2015, Spinelle2017, Maag2018}, whereas some studies report the opposite~\cite{Cordero2018}.
This might be explained by the fact that the relationship between the pollutants and how they vary in time is a very complex function.
Therefore, the distribution of pollutants and characteristics of environmental variables can influence whether \ac{LR} and \ac{MLR} models are sufficient for approximating the underlying relationships in data.

The complexity of the relationships that can be captured with ANNs depends on the structure of the network.
Standard FFNN structures mostly have been designed for capturing a non-linear function between input and output.
However, more complex structures that can incorporate additional considerations as part of the network, such as temporal structure or feature learning, have been proposed.
As an example, \acp{RNN} are a class of \acp{ANN} designed for incorporating temporal structure and thus well-suited for modeling time series.
In an \ac{RNN}, input values or neuron outputs of the previous time step can influence the state of the \ac{ANN} in the current time step.
For the calibration of low-cost air quality measurements, Esposito et al.~\cite{Esposito2016} tested two more complex \ac{RNN} architectures: \ac{TDNN} and \ac{NARX}.
They report that \acp{NARX} have a better performance than \acp{FFNN}, and \acp{TDNN} achieve the best performance, which might be explained by the fact that \acp{RNN} take into consideration previous time steps, therefore encoding input value changes over time.

It is also possible to build hybrid models by using layers of different architectures.
For example, we can use an \ac{RNN} layer to learn relationships between different points in the input data, an \ac{FFNN} layer to learn another level of abstraction, and a final \ac{FFNN} layer to constrain the size of the output.
These hybrid models are called \ac{DL} models.
The idea behind this is to create a model that can learn multiple levels of abstraction of the input data \cite{LeCun2015}.

\begin{description}
    \item[Advantages:] Very flexible learner that can approximate any function, given enough layers and neurons.
    It can automatically learn the relationships between the features and learn multiple response values simultaneously.
    \item[Disadvantages:] Its complexity makes it extremely expensive in computing resources to train, requiring dedicated hardware for doing so quickly.
    It also needs a huge amount of data for avoiding overfitting.
    \acp{ANN} are typically black-box algorithms.
\end{description}

\subsection{Gaussian Processes}

A \ac{GP} is a model that combines multiple Gaussian random variables into a joint distribution to estimate the function that models the data.
It is a non-parametric approach, which means that there is no need to specify the number of parameters.
However, similarly to \ac{SVR}, \acp{GP} need a kernel function to be specified.
Detailed information about \acp{GP} and how to train them can be found in Rasmussen et al.
\cite{Rasmussen2004}.

\acp{GP} make weak assumptions on the distribution of input data, and hence are well suited for calibrating low-cost air quality sensors.
Another benefit of \acp{GP} is that it is possible to plot the probability distributions used to model data.
\acp{GP} may require a training dataset that is larger than linear models and \acp{SVM} would require, but generally smaller than more complex models such as \acp{ANN}.
\acp{GP} are lazy learners, meaning they do not need to be trained, but instead, they approximate the function of the training dataset while predicting.
However, this means that the whole training dataset needs to be kept in memory, and every time that it needs to predict the target variable, it needs to compare it to the probability distribution of the features.
This means that \acp{GP} have high memory complexity, which might be unpractical if the deployed \ac{AQS} units do not communicate with a central infrastructure and need to run calibration in the field.
\acp{GP} have been used in the context of low-cost \ac{AQS} calibration by Cheng et al.~\cite{Cheng2014} and Gao et al.~\cite{Gao2016} on top of \ac{FFNN} to improve its performance.
In both studies, \acp{GP} were able to improve calibration performance compared to a standalone \ac{FFNN}.

\begin{description}
    \item[Advantages:] Non-parametric model, no need to specify the number of parameters except for the kernel.
    Makes only weak assumptions on the distribution of data.
    Gives probability distribution for the predictions.
    \item[Disadvantages:] High memory complexity, requires a kernel function and can be sensitive to parameters of the kernel function.
\end{description}

\subsection{Other Machine Learning Paradigms}

Most of the early work on air quality sensor calibration focused on validating and testing traditional machine learning algorithms, as discussed in the previous sections.
Recently the focus has shifted towards more advanced paradigms that address specific challenges in the development of calibration systems.
Below we briefly discuss some of the most important ones:

\bparagraph{Transfer learning.} It may happen that a model trained in one domain (e.g., at a specific time interval, at a specific location, or for specific sensor hardware) is not accurate in another domain and the machine learning model has to be trained separately for each domain.
The idea in {\em transfer learning} is to mitigate this problem and to make it possible to use commonalities between the different domains to train better and more accurate models, with a smaller amount of data needed per domain~\cite{zhang2018crosssense, Zhuang2020}.
Transfer learning can transform a pre-trained model so that it can be applied to a different domain.
While parameters of the model need to be tuned for the new environment, the technique allows adapting previous solutions to new problems using existing models and data.

\bparagraph{Speeding Up Training.} Several machine learning paradigms are closely related to transfer learning, but instead of attempting to generalize to other domains, focus on making the learning process easier.
Firstly,  {\em meta learning}, also known as {\em learning to learn}, focuses on systematically observing how machine learning algorithms learn on different domains and using this knowledge to speed up the learning process on a new domain~\cite{Vanchoren2018}.
Another related paradigm is {\em few-shot learning}, where a collection of techniques is used to generalize to new tasks with only a small amount of labeled training samples~\cite{Wang2020}.
{\em Semi-supervised learning} focuses on using a limited amount of labeled ground truth measurements to support training ~\cite{vanengelen2020}.
Semi-supervised learning algorithms utilize the structures found in the measurements with and without the ground truth to make better models than would be possible with the ground truth measurements alone.

\bparagraph{{Incremental Learning.}} Calibration models suffer accuracy loss in long-term field deployments due to many negative effects, such as sensor drift, changes in the probability distribution of priors due to seasonal changes, and so on.
Hence, it is important to have a more accurate model by updating the original model, taking into consideration new measurements and changes.
Incremental learning is a continuous learning approach where the learning process takes place whenever new samples emerge, and it adjusts what has been learned according to the new samples.
Continuous learning, online learning, and adaptive learning are quite similar machine learning paradigms to incremental learning.
De Vito et al.~\cite{vito2020} show that adaptive and incremental strategies, such as performing periodic recalibration, can improve the performance of initially trained calibration models for low-cost sensors.
Through experiments carried out on measurements collected from 18-months electrochemical sensors deployments, monitoring CO, NO\textsubscript{2}, and O\textsubscript{3} concentrations, the authors demonstrate that such strategies improve the overall performance of calibration models and make them less sensitive to seasonal changes or other variations.

\bparagraph{Dimensionality Reduction.} When a large number of variables are used for learning the calibration function, the complexity of the measurements and the resulting models can be reduced using dimensionality reduction techniques.
Dimensionality reduction is a special case of {\em unsupervised learning}, and can be implemented using methods such as cluster analysis~\cite{Hastie2009}.

\subsection{Network Calibration Strategies}

In practice, individual sensors rarely have continuous access to reference measurements or even to the same source of reference measurements~\cite{motlagh2020towards}.
\textit{Network calibration} refers to the process of post-deployment calibration, which can either focus on establishing a calibration function for a new sensor with the help of existing sensors or re-training the model when concept drift occurs.
Below we briefly cover the main techniques for network calibration, and we refer to the survey by Maag et al.~\cite{Maag2018survey} for a more detailed discussion on network (re-)calibration strategies. 

\bparagraph{Blind Calibration.} This technique attempts to adapt calibration models so that there is a high similarity between measurements across the entire measurement network.
Blind calibration typically assumes that the deployment is dense and neighboring sensors have closely matching or at least highly correlated measurements.
Balzano et al.~\cite{Balzano2007} alleviate the deployment density requirement but require measurements to be correlated across the network.
Unfortunately, this is not the situation in the air quality monitoring system, as we have mentioned earlier that the measurements of sensors cannot be guaranteed to be identical, and even a single city block can witness significant variations in pollutant concentrations.
Tsujita et al.~\cite{Tsujita2005} developed a gas sensor system that incorporates an auto-calibration method.
They developed a case study of \ch{NO2} distribution, performed by continuously installing a low-cost \ch{NO2} sensor in different locations where there is no professional-grade reference station in the vicinity.
Sensors were then calibrated to the reference station when the \ch{NO2} concentrations were low and expected to have almost identical values in deployed areas.
Such approach is only applicable in areas where the pollution level regularly reaches a low level.
Blind calibration usually operates under the assumption that the distribution of the concentration of pollutants is uniform in a certain region~\cite{Moltchanov2015, Mueller2017}, and is only useful for offset and gain calibration~\cite{cheng2019ict}.
Miskell et al. \cite{Miskell2019} designed a hierarchical network framework incorporating a proxy model, a measurement model, and a semi-blind calibration model.
They adjust the gain and offset of a low-cost sensor's measurements to the first two moments of the probability distribution of the nearest reference station, and use the adjusted data as a proxy for most sites.
The developed algorithms achieve great success in detecting and correcting sensor drift, and the designed framework can deliver reliable high temporal-resolution ozone data at neighborhood scales.

\bparagraph{Opportunistic and Collaborative Calibration.} When the sensors are mobile, opportunistic encounters between devices can be used to create virtual reference points that can be utilized for calibration~\cite{Xiang2012}.
Saukh et al.~\cite{Saukh2015} propose a PM calibration approach that uses opportunistic encounters to obtain reference sensor measurements.
Their approach uses these measurements in a multi-sensor data fusion approach to learn the calibration function.
Opportunistic calibration can either rely on measurements from a single device or use a \textit{collaborative} approach, where devices share reference measurements and use them to learn the calibration function~\cite{motlagh2020towards}.
A further extension is \textit{multi-hop calibration}, which leverages encounter-based opportunistic calibration to propagate the calibration to other sensors.
Specifically, newly calibrated sensors are considered as references for devices that cannot directly access reference devices, and thus calibration parameters and measurements are recursively propagated throughout the measurement network.
As an example, Maag et al. \cite{Maag2017} propose a multi-hop calibration technique, sensor array network calibration (SCAN), for dependent low-cost sensors.
SCAN minimizes error accumulation over sensor arrays and has been theoretically proven to be free from regression dilution. 

\bparagraph{Calibration Transfer.} This technique refers to the use of transfer learning to adapt calibration functions across domains.
Calibration transfer has been mainly performed in-lab for pre-deployment calibrating electronic noses to reduce calibration overhead in mass production where each device needs to be calibrated due to inter-device difference~\cite{Yan2016, Yan2018}.
During the calibration process, a pre-trained calibration model with fine-tuning parameters is produced and later adopted on the target sensor.
Cheng et al.~\cite{cheng2019ict} propose an in-field calibration transfer with a large-scale real-world PM monitoring deployment. The authors assume the target location holds a similar distribution of the ground truth compared to the source location, and that the required transformation is approximately linear.
Cheng et al.~\cite{Cheng2020} recently designed an air quality map generation scheme named MapTransfer.
Their main idea is to enlarge the current sensor measurements in a downscaled sparse deployment with suitable historical data from a shorter duration dense deployment.
A learning-based data selection scheme is adopted to select the best matching data, and a multi-output Gaussian process model is used for fusing the best-selected data with the current measurement.
In the experiment, they consider data spanning the whole year 2018 from 200 sensors as the dense deployment, and data spanning half of the year 2019 from 50 randomly selected sensors out of 220 as the sparse deployment.
These calibration techniques are very powerful for dealing with post-deployment scenarios, where irregular or even no access to reference measurements can be a common issue.

%% file: tables/cal_summary.tex
\begin{table*}
    \centering
    \caption{Summary of calibration studies.}
    \label{tab:cal_summary}
    \resizebox{\linewidth}{!}{
    \begin{tabular}{c l c c c | c | c | c}
        \toprule
        ML model & Reference & Training data & Test data & Temporal & Mobility & Exploits & Online \\
         & & length & length & resolution &  & temporality & training \\
        \midrule
        \multirow{3}{*}{LR}
         & Hasenfratz et al. \cite{Hasenfratz2012} & NR & $\leq$ 2 mos. & 5 s & X & & \\
         & Lin et al. \cite{Lin2015} & NR & $\leq$ 2 mos. & 5 mins & & & \\
         & Saukh et al. \cite{Saukh2015} & NR & $\leq$ 6 mos. & 10 s, 20 s & X & & \\
        \hline
        \multirow{5}{*}{MLR}
         & Maag et al. (2016) \cite{Maag2016} & 4 wks. & 1.25 yrs. & 5 s & X & & \\
         & Maag et al. (2018) \cite{Maag2018} & 2.1 days & 2.7 wks. & NR & X & & \\
         & Liu et al. \cite{Liu2017} & NR & 36 h & 1 min & X & & \\
         & Cordero et al. \cite{Cordero2018} & NR & $\leq$ 30 days & NR & & & \\
         & Zimmerman et al. \cite{Zimmerman2018} & NR & 1.4--15 wks. & 15 mins & & & \\
        \hline
        SVM & Cordero et al. \cite{Cordero2018} & NR & $\leq$ 30 days & NR & & & \\
        \hline
        \multirow{3}{*}{RF}
         & Borrego et al. \cite{Borrego2016, Borrego2018} & 12.6 days & 1.4 days & 1 min/1 h & & & \\
         & Cordero et al. \cite{Cordero2018} & NR & $\leq$ 1 mo. & NR & & & \\
         & Zimmerman et al. \cite{Zimmerman2018} & 5.6 days & 1.4--15 wks. & 15 mins & & & \\
        \hline
        \multirow{7}{*}{FFNN}
         & DeVito et al. (2008) \cite{DeVito2008} & 8 mos. & 3 mos. & 1 h & & & \\
         & DeVito et al. (2009) \cite{DeVito2009} & 2 wks. & 7 mos. & 1 h & & & \\
         & Spinelle et al. \cite{Spinelle2015, Spinelle2017} & 1 wk. & 4.3 mos. & 1 min & & & \\
         & Esposito et al. \cite{Esposito2016} & 1 wk. & 3 wks. & 1 min & & X & \\
         & Borrego et al. \cite{Borrego2016, Borrego2018} & 12.6 days & 1.4 days & 1 min/1 h & & & \\
         & Maag et al. \cite{Maag2018} & 2.1 days & 2.7 wks. & NR & X & & \\
         & Cordero et al. \cite{Cordero2018} & NR & $\leq$ 30 days & NR & & & \\
        \hline
        \multirow{2}{*}{FFNN + GP}
         & Cheng et al. \cite{Cheng2014} & 3.5 mos. & 2 mos. & 5 mins & X & & \\
         & Gao et al. \cite{Gao2016} & NR & NR & 1 min & X & & \\
        \hline
        NARX & Esposito et al. \cite{Esposito2016} & 1 wk. & 3 wks. & 1 min & & X & \\
        \hline
        TDNN & Esposito et al. \cite{Esposito2016} & 1 wk. & 3 wks. & 1 min & & X & \\
        \bottomrule
    \end{tabular}
    }
\end{table*}

%% file: sections/performance.tex
\section{Measuring performance and comparing models}
\label{sec:performance}

Machine-learning-based calibration models are only useful if they can consistently improve the accuracy of the sensor measurements produced by a low-cost sensing unit.
To ensure this indeed is the case, calibration models need to be validated against measurements collected from high-cost reference stations.
Next, we discuss validation methods and how they can be applied to low-cost calibration, and compare existing low-cost air quality calibration studies by selecting the most commonly used performance measures in them.

\subsection{Performance Measures}

The performance of calibration models is typically expressed through one or more performance measures, which are functions that characterize the dissimilarity between the output of the calibration model and the ground truth values obtained from a reference station.
Existing studies have used differing performance measures, which makes it difficult to compare performance across studies.
In the following, we briefly summarize the main performance measures.

\bparagraph{Absolute Error Measures.} \ac{MSE} is the standard error measure for assessing the performance of regression models.
MSE is defined as:
\begin{equation}
    \text{MSE} = \frac{1}{m} \sum_{i = 1}^m (y_i - \hat{y}_i)^2
\end{equation}
where $m$ is the number of samples, $\hat{y}_i$ is the predicted value and $y_i$ is the actual value of a sample.
\ac{MSE} is useful to evaluate the performance during training and to define a cost function because of its simplicity.
A variation of \ac{MSE} is \ac{RMSE}, defined as:
\begin{equation}
    \text{RMSE} = \sqrt{\frac{1}{m} \sum_{i = 1}^m (y_i - \hat{y}_i)^2}
\end{equation}
Another absolute error measure is \ac{MAE} which is defined as:
\begin{equation}
    \text{MAE} = \frac{1}{m} \sum_{i = 1}^m \left| y_i - \hat{y}_i \right|
\end{equation}
Finally, \ac{MBE} is defined as:
\begin{equation}
    \text{MBE} = \frac{1}{m} \sum_{i = 1}^m{\left(y_i - \hat{y}_i\right)}.
\end{equation}

MSE and RMSE are very similar.
MSE can be interpreted geometrically as the average fit of points to a regression model, whereas RMSE is the average distance of points from the regression model.
RMSE and MSE weigh errors proportionally to their magnitude, whereas MAE weighs all errors equally.
This makes RMSE and MSE more sensitive to outliers~\cite{Chai2014}, suggesting that MAE is a better measure for measuring the average performance, whereas (R)MSE is useful for measuring a model's sensitivity to outliers.
In practice, it is recommended to use both measures together as this provides complementary information on the model's performance~\cite{Chai2014}.
MBE, on the other hand, measures whether the average error is positive or negative and can be used to determine whether the model underestimates or overestimates the pollutant values.
Spinelle et al.
\cite{Spinelle2015, Spinelle2017} divide \ac{RMSE} and \ac{MBE} by 
the standard deviation of the reference measurements and combine the resulting values into a target diagram.
Target diagrams are useful for visualizing these two performance metrics and for quickly comparing different models.

\bparagraph{Relative Error Measures.} The alternative to absolute measures is to rely upon a relative error measure which expresses the error proportionally to the true measurements.
The most popular relative error measure is \ac{MRE} which is defined as: 
\begin{equation}
    \text{MRE} = \frac{1}{m} \sum_{i = 1}^m \left| \frac{y_i - \hat{y}_i}{y_i} \right|
\end{equation}
A related measure is \ac{MAPE} which expresses \ac{MRE} as a percentage:
\begin{equation}
    \text{MAPE} = \frac{1}{m} \sum_{i = 1}^m \left| \frac{y_i - \hat{y}_i}{y_i} \right| \cdot 100 \%
\end{equation}
\ac{MRE} is useful for expressing how far estimated values are from the reference values, whereas MAPE is useful for characterizing performance when the same model is applied for multiple pollutants.

\bparagraph{Coefficient of Determination.} The coefficient of determination, or $R^2$, measures how much a variable influences another variable.
For a calibration model, $R^2$ measures the percentage of variance that the model explains.
To compute $R^2$ we need to compute two variability measures, namely the total sum of squares:
\begin{align}
\text{SS}_{\text{tot}} = \frac 1m \sum_{i = 1}^m (y_i - \Bar{y})^2 
\end{align}
and the sum of squares of residuals:
\begin{align}
\text{SS}_{\text{res}} = \frac 1m \sum_{i = 1}^m (y_i - \hat{y}_i)^2 = \text{MSE}.
\end{align}
Here $\Bar{y}$ is the mean of the target data.
The $R^2$ is now given as:
\begin{align}
R^2 = 1 - \frac{\text{SS}_{\text{res}}}{\text{SS}_{\text{tot}}} 
\end{align} 
$R^2$ can be useful in the low-cost \ac{AQS} calibration scenario to assess how closely the distribution of the predicted values matches the distribution of the ground truth measurements.

\bparagraph{Uncertainty Measures.} Air quality standards typically associate bounds on the uncertainty that the measurements can contain.
From an algorithmic point-of-view, uncertainty can be considered as a measure of \textit{robustness} as it provides insights into the operating bounds of the calibration framework.
As an example, the Clean Air directive of the European Union assigns maximum deviations for the $95\%$ uncertainty of measurements, with the precise deviation depending on the pollutant~\cite{eu-2008-50}.
Uncertainty is typically measured using the standard deviation of measurements, which specifies the \textit{standard uncertainty} of the measurements.
Alternatively, the \textit{relative standard uncertainty} specifies measurement uncertainty relative to the magnitude of the measurements.
The standard deviation of measurements can be interpreted as a confidence interval of $68\%$ which leaves a broad margin for the measurements to deviate.
A tighter bound can be obtained using \textit{expanded uncertainty} which is defined as the standard deviation multiplied by a \textit{coverage factor} $k$ which determines the bounds of the uncertainty region.
A coverage factor of $k=2$ roughly translates into a confidence interval of $95\%$ whereas a coverage factor of $k=3$ corresponds to a confidence interval of $99\%$.
Similarly to the standard uncertainty, \textit{expanded relative uncertainty} is a measurement of uncertainty defined relative to the magnitude of the measurements.
In practice, the robustness of the evaluation setup of air quality models is measured using expanded relative uncertainty, as that allows to compare uncertainty across areas with differing pollutant concentrations.

\bparagraph{Best Practices.} \ac{MSE}, \ac{RMSE} and $R^2$ are closely related.
This means that, if we rank some models according to one or the other measure separately, the ranking positions for the models in both rankings will be the same.
The same holds for \ac{MRE} and \ac{MAPE}.
However, measures that are not directly related, such as \ac{RMSE}, \ac{MAE}, and \ac{MRE}, do not necessarily result in the same ranking.
Hence, in practice, the recommended approach is to use multiple performance measures and take into account how they are affected by the properties of the data.
Visual aids, such as target diagrams, should also be used so that different performance measures can be visually compared.
A summary of the performance measures can be found in \autoref{tab:measures-comparison}.

\input{tables/measures-comparison.tex}

\subsection{Evaluation Criteria for Low-Cost Deployments}

To evaluate the suitability of air quality sensor calibration methods considered in this survey for long-term deployments of low-cost sensors, we classify them based on four evaluation criteria.
\begin{itemize}
    \item The robustness of the evaluation setup of the method, based on the length of the test dataset
    \item The resolution, determined by the length of the smallest temporal step modeled
    \item The accuracy of the method as reported by the authors
    \item The scalability of the sensor technology, emphasizing technologies are low-cost and that can operate independently for long periods
\end{itemize}
Each method is classified as \textit{Low, Medium} or \textit{High} with respect to each of the four criteria.
Methods that could not be classified based on the information available on it are marked with \textit{N/A}.
The rest of this section explains how the criteria are determined for a given method.

\bparagraph{Robustness.} This criterion is based on the variability of the measurements, and it is indirectly linked to the length of the test dataset. Longer test datasets have a higher probability to include seasonal and weekly variations, which result in a wider spectrum of pollution and environmental values in the data. This in turn ensures the evaluation setup considers the robustness of the model in the presence of such variations---and consequently can provide better estimates of uncertainty. 
Methods with one-month long datasets or shorter are classified \textit{Low} in terms of Robustness, and one-year-long datasets or longer are classified as having \textit{High} robustness. Note that optimal test data would include multiple years, so the effects of variations between years in the model can be mitigated. In practice, collecting such long-term data with an identical measurement setup is often very difficult due to practical constraints, sensor failures, and other sources of errors, such as sensitivity drift of the measurement hardware.

\bparagraph{Resolution.} This score is based on the temporal resolution of the data used in a model.
Methods using data with a resolution of one hour or coarser are classified \textit{Low} and those using data with a resolution of one minute or finer are classified \textit{High}. We note that the main purpose of this classification is to consider the suitability of the technologies for near real-time services, such as estimation of current air pollution levels and calculation of the relevant air quality index value.
In some applications, such as long-term assessment of pollution levels or prediction of hazardous areas, a one-hour resolution is sufficient.
Hence, classification is not intended as an assessment of quality, but rather as a statement on the level of granularity that is offered. 

\bparagraph{Accuracy.} This score is based on the accuracy of the model.
This is the most complex to estimate since different studies use different evaluation measures.
We define it as follows.
If a study uses RMSE, or as an alternative, MAE, it is used as the base value of the performance of the models.
Since Spinelle et al.~\cite{Spinelle2015, Spinelle2017} do not provide an exact value, we use a conservative estimate of MAE obtained by analyzing the residual plots that they provide.
Models of studies that do not use comparable similarity measures are marked \textit{N/A}.
We group the models by the type of pollutant they predict.
In each group, the model with the best performance is classified \textit{High}, and the model with the worst performance is classified \textit{Low}. Other models are classified as having \textit{Medium} Accuracy.

\bparagraph{Scalability.} This criterion is based on the sensor technologies used to produce the input data of a model.
The score is determined based on the suitability of such sensor technology for large-scale deployments, taking into account the typical cost, robustness, and lifespan of such technology.
\ac{DiSC} sensors are classified \textit{Low} due to their higher cost, \acp{NDIR}, \acp{PID} and \acp{OPC} are classified \textit{Medium}, while \ac{LSP}, \ac{MOS} and \ac{EC} sensors are classified \textit{High} (higher value for scalability is indicative of the technology being a better fit for large-scale deployments).
The Scalability classification of each method is computed as the mean of the classifications of the sensors used in the method. For example, a method with one \ac{MOS}, one \ac{EC} and one \ac{NDIR} sensor is classified \textit{High}, while a method with one \ac{MOS} and one \ac{NDIR} sensor is classified \textit{Medium}.

\input{tables/appcomparison_new.tex}

\subsection{Comparing Studies}
\label{ssec:comparing}

The different studies surveyed for this article are compared and summarized in Table~\ref{tab:appcomparison}.
Below we separately discuss the studies according to each of the four criteria.

\bparagraph{Robustness.} The study by Maag et al.~\cite{Maag2016} is the only one scoring high for robustness. 
This is because they use a dataset that spans more than a year to test the model, which is by far the longest among all studies surveyed for this article.
The next highest is the study by De Vito et al.
(2009)~\cite{DeVito2009}, which uses a test dataset spanning about 7 months.
From the comparison, we can observe that most studies use relatively short test datasets, which are unlikely to capture the full extent of seasonal variations.
Most datasets used in the literature are proprietary and specific to a single deployment.
Taken together, these factors mean that an objective comparison of calibration methods is currently highly difficult due to short measurement periods and a lack of openly available reference datasets.

\bparagraph{Resolution.} Many studies score high in Resolution because they construct models using data with a temporal resolution lower or equal than 1 minute.
These studies are Maag et al.
(2016) \cite{Maag2016}, Spinelle et al.
(2015, 2017) \cite{Spinelle2015, Spinelle2017}, Cheng et al.
\cite{Cheng2014}, Esposito et al.
\cite{Esposito2016}, Liu et al.
\cite{Liu2017a}, Saukh et al.
\cite{Saukh2015}, Hasenfratz et al.
(2012) \cite{Hasenfratz2012}, and Gao et al.
\cite{Gao2016}.
Other studies with a good resolution score are Lin et al.
\cite{Lin2015} and Zimmerman et al.
\cite{Zimmerman2018}.
Every other study uses data with a 1-hour resolution or does not report the used resolution.
These result in Resolution scores of Low and N/A, respectively.
This suggests that most studies use data with a good temporal resolution.
Note that we only compared temporal resolution, and the picture would be completely different if we considered also the spatial resolution of the studies.
While air quality deployments are increasingly commonplace, practically all studies only consider measurements from a single geographic location.
Together with the lack of openly available datasets, this means that it is difficult to currently assess how well the methods perform across spatial variations -- let alone considering the combined effect of spatial and temporal variations.

\bparagraph{Accuracy.} The \ac{RF} model by Zimmerman et al.~\cite{Zimmerman2018} has the top ranking in accuracy for most pollutants, namely \ch{CO}, \ch{CO2}, \ch{NO2}, and \ch{O3}.
Notable mentions with good performance are the MLR model by Maag et al.~\cite{Maag2016} for \ch{CO}, the FFNN model by Spinelle et al.~\cite{Spinelle2017} for \ch{CO2}, the TDNN and NARX models by Esposito et al.~\cite{Esposito2016} for \ch{NO2}, and the RF model by Borrego et al.~\cite{Borrego2018} for \ch{NO2} and \ch{O3}.
Only a few models have been developed for the rest of the pollutants, and a few of the studies that present them do not present meaningful performance measures.
Because of these reasons, we will only mention the best model for each.
For \ch{NO}, the best model is the FFNN model by Spinelle et al.
(2017) \cite{Spinelle2017}, for \ch{NO_x}, the best is the TDNN model by Esposito et al.~\cite{Esposito2016}, for \ch{SO2}, the best model is the RF model by Borrego et al.
(2018)~\cite{Borrego2018}, and for For \ch{PM_{2.5}}, the best is the FFNN model with GP by Cheng et al.~\cite{Cheng2014}.
It is noteworthy that the best performing model for almost all pollutants is different (with the sole exception being the FFNN model that is best for two pollutants).
Whether this is due to a lack of sufficiently generalizable models or differences in pollutant characteristics is currently an open issue that warrants further investigation.
Indeed, the target and input variables of the models tend to vary across studies, which makes it difficult to draw any generalizable conclusions about the types of models that are needed to support different pollutants.
The comparison is further exacerbated by the fact that the sensor technology used in each study tends to differ from the other studies.

\bparagraph{Scalability.} The Scalability score evaluates the sensor technologies used in the studies.
The studies with the highest score are those by De Vito et al.~\cite{DeVito2009}, Hasenfratz et al.~\cite{Hasenfratz2012}, Lin et al.~\cite{Lin2015}, Saukh et al.~\cite{Saukh2015}, Maag et al.~\cite{Maag2018} and Zimmerman et al.~\cite{Zimmerman2018}.
All of these studies have in common the fact that they use \ac{MOS} sensors -- which, along with \ac{EC} sensors, seem to be the easiest and cheapest to integrate into large-scale deployments.
Studies with \ac{EC} sensors were also ranked \textit{High}.
Other studies with a good score are Cheng et al.~\cite{Cheng2014}, Gao et al.~\cite{Gao2016}, and Liu et al.~\cite{Liu2017a}, which all use \ac{LSP} sensors.
The rest of the studies have a score that ranges from middle to low, 
since they use a combination of sensors including also high-cost sensors,
such as Zimmerman et al.~\cite{Zimmerman2018} which uses \ac{NDIR}.
Studies with a higher than average score are Spinelle et al.~\cite{Spinelle2015, Spinelle2017}, and Zimmerman et al.~\cite{Zimmerman2018}.
Scalability is an essential metric for assessing the suitability of technologies for large-scale deployments and a useful tool for fostering the adoption of denser sensor deployments.
Currently comparing scalability is difficult due to a lack of suitable points of reference.
The cost of low-cost sensors will always be several orders of magnitude lower than that of professional reference stations, but a combination of both is required to ensure the low-cost sensors can be calibrated.
To measure scalability, suitable economic and geographic references are needed.
As an example of this, Motlagh et al.~\cite{motlagh2020towards} estimated a cost for a dense deployment of sensors in central Beijing.
Having such geographic estimates that can relate resolution of monitoring, cost of deployment and improvements in accuracy, would improve comparison of sensor units and their suitability for large-scale deployments.
Naturally, these reference points should be sufficiently varied, as population density is another factor that guides the requirements for the density of deployments.

%% file: tables/measures-comparison.tex
\begin{table*}
    \centering
    \caption{Comparison of common performance measures.}
    \label{tab:measures-comparison}
    \resizebox{\linewidth}{!}{
    \begin{tabular}{ l  c  >{\raggedright\arraybackslash}m{10em}  >{\raggedright\arraybackslash}m{14em}  >{\raggedright\arraybackslash}m{16em} }
        \toprule
        Method & Metric & Formula & Advantages & Disadvantages \\
        \midrule
        MSE & No &
        $ \frac{1}{m} \sum_{i = 1}^m (y_i - \hat{y}_i)^2 $ &
        Simple measure, can be used as a cost function. Useful for measuring the model's sensitivity to outliers. &
        Tends to exaggerate errors, especially with noisy data. For very clean data it might overestimate the model performance. \\
        \hline
        RMSE & Yes &
        $ \sqrt{\frac{1}{m} \sum_{i = 1}^m (y_i - \hat{y}_i)^2} $ &
        Same as MSE, but in the same dimension as the target values. &
        Same as MSE. \\
        \hline
        MAE & Yes &
        $ \frac{1}{m} \sum_{i = 1}^m \left| y_i - \hat{y}_i \right| $ &
        Useful for measuring the "average" performance of a model. & 
        Underestimates the outliers. \\
        \hline
        MBE & No &
        $ \frac{1}{m} \sum_{i = 1}^m{\left(y_i - \hat{y}_i\right)} $ &
        Useful for measuring the bias, and to see to which value the average error tends. &
        It takes into account the bias only. It can't be used to evaluate the actual performance of a model. \\
        \hline MRE & Yes &
        $ \frac{1}{m} \sum_{i = 1}^m \left| \frac{y_i - \hat{y}_i}{y_i} \right| $ &
        Useful for expressing the average error in proportion to the target values. &
        Tends to exaggerate the error for small values, and underestimate the error for big values. \\
        \hline
        MAPE & Yes & 
        $ \frac{1}{m} \sum_{i = 1}^m \left| \frac{y_i - \hat{y}_i}{y_i} \right| \cdot 100 \% $ & Same as above. & Same as above. \\
        \hline
        $R^2$ & No &
        $ 1 - \frac{\text{SS}_{\text{res}}}{\text{SS}_{\text{tot}}} $ &
        Useful for measuring how much the variance is accounted for by the model. &
        Same as MSE and RMSE. \\
        \bottomrule
    \end{tabular}
    }
\end{table*}

%% file: tables/appcomparison_new.tex
\begin{sidewaystable*}
    \centering
    \caption{Comparison of the different machine learning approaches for low-cost \acp{AQS} calibration, including performance measures and our evaluation criteria. For each pollutant, rows are ranked using our evaluation criteria-based final ranking in descending order.}
    \label{tab:appcomparison}
    \resizebox{\linewidth}{!}{\begin{tabular}{c | c | l | c c c c c c c c c c | c c c c | c c c c | c c c c }
        \toprule
        Target & Method & Reference & \multicolumn{10}{ c|}{Features} & \multicolumn{4}{ c|}{Atmospheric features} & \multicolumn{4}{ c| }{Performance measures} & \multicolumn{4}{c}{Evaluation Criteria} \\
         & & & CO & \ac{CO2} & NO & \ac{NOx} & \ac{NO2} & \ac{O3} & \ac{SO2} & VOCs & PM\textsubscript{2.5} & PM\textsubscript{10} & T & RH & AH & WS & RMSE & MAE & MAPE & $R^2$ & 
         Robustness & Resolution & Accuracy & Scalability \\
        \midrule
        \multirow{8}{*}{CO} & MLR & \cite{Maag2016} & x &  &  &  & x & x &  &  &  &  & x & x &  &  & 0.048 ppm & NR & NR & NR & 
        High & High & Medium & Medium  \\
         & RF & \cite{Zimmerman2018} & x & x &  &  & x & x &  &  &  &  & x & x &  &  & NR & 7.9 ppb & NR & 0.99 & 
         Medium & Medium & High & Medium  \\
         & FFNN & \cite{Spinelle2017} & x & x &  &  & x &  &  &  &  &  & x &  & x &  & NR & 0.05--0.12 & NR & 0.367 & 
         Medium & High & Medium & High \\
         & MLR & \cite{Zimmerman2018} & x &  &  &  &  &  &  &  &  &  & x & x &  &  & NR & 39 ppb & NR & 0.94 & 
         Medium & Medium & Medium & High  \\
         & FFNN & \cite{DeVito2009} & x &  &  &  &  &  &  & x &  &  &  &  &  &  & NR & 0.31 ppm & 27.00 \% & NR & 
         Medium & Low & Low & High  \\
         & RF & \cite{Borrego2018} & x & x & x &  & x & x & x &  &  & x & x & x &  &  & NR & 0.07 ppm & 280.00 \% & 0.88 & 
         Low & Low & Medium & Medium  \\
         & FFNN & \cite{Borrego2018} & x & x & x &  & x & x & x &  &  & x & x & x &  &  & NR & 0.09 ppm & 34.00 \% & 0.51 & 
         Low & Low & Medium & Medium \\
         & LR & \cite{Saukh2015} & x &  &  &  &  &  &  &  &  &  &  &  &  &  & 0.23 ppm & NR & NR & NR & 
         Low & High & Medium & High  \\
        \hline
        \multirow{5}{*}{CO\textsubscript{2}} & FFNN & \cite{Spinelle2017} & x & x &  &  & x &  &  &  &  &  &  &  &  &  & NR & 2--18 & NR & 0.787 & 
        Medium & High & Medium & Medium \\
         & RF & \cite{Zimmerman2018} & x & x &  &  & x & x &  &  &  &  & x & x &  &  & NR & 1.7 ppb & NR & 0.99 & 
         Medium & Medium & High & Medium \\
         & MLR & \cite{Zimmerman2018} &  & x &  &  &  &  &  &  &  &  & x & x &  &  & NR & 19 ppb & NR & 0.49 & 
         Medium & Medium & Medium & Medium  \\
         & FFNN & \cite{Maag2018} &  &  &  &  &  & x &  & x &  &  & x &  &  &  & NR & 44 ppb & NR & 0.94 & 
         Medium & N/A & Medium & High   \\
         & MLR & \cite{Maag2018} &  &  &  &  &  & x &  & x &  &  & x &  &  &  & NR & 135 ppb & NR & 0.49 & 
         Medium & N/A & Low & High  \\
        \hline
        \multirow{3}{*}{NO} & FFNN & \cite{Spinelle2017} &  &  & x &  & x &  &  &  &  &  & x & x &  &  & NR & 0.1--0.5 & NR & 0.208 & 
        Medium & High & High & Medium \\
         & FFNN & \cite{Borrego2018} & x & x & x &  & x & x & x &  &  & x & x & x &  &  & NR & 2.39 ppb & 84.00 \% & 0.39 & 
         Low & Low & Medium & Medium \\
         & RF & \cite{Borrego2018} & x & x & x &  & x & x & x &  &  & x & x & x &  &  & NR & 4.76 ppb & 127.00 \% & 0.74 & 
         Low & Low & Low & Medium \\
        \hline
        \multirow{15}{*}{NO\textsubscript{2}} & MLR & \cite{Maag2016} & x &  &  &  & x & x &  &  &  &  & x & x &  &  & 5.13 ppb & NR & NR & NR & 
        High & High & Medium & Medium \\
         & RF & \cite{Zimmerman2018} & x & x &  &  & x & x &  &  &  &  & x & x &  &  & NR & 0.5 ppb & NR & 0.99 & 
         Medium & Medium & High & Medium  \\
         & FFNN & \cite{Spinelle2015} & x &  &  &  & x & x &  &  &  &  &  &  & x &  & NR & 2--4.5 ppb & NR & 0.596 & 
         Medium & High & Medium & Medium  \\
         & FFNN & \cite{DeVito2009} & x &  &  & x & x & x &  & x &  &  &  &  &  &  & NR & 10.1 ppb & 22.00 \% & NR & 
         Medium & Low & Low & High \\
         & MLR & \cite{Zimmerman2018} &  &  &  &  & x &  &  &  &  &  & x & x &  &  & NR & 4.6 ppb & NR & 0.59 & 
         Medium & Medium & Medium & High  \\
         & TDNN & \cite{Esposito2016} &  &  & x &  & x & x &  &  &  &  & x & x &  &  & NR & 1.27 ppb & 22.00 \% & NR & 
         Medium & High & Medium & High  \\
         & NARX & \cite{Esposito2016} &  &  & x &  & x & x &  &  &  &  & x & x &  &  & NR & 1.30 ppb & 21.00 \% & NR & 
         Low & High & Medium & High  \\
         & FFNN & \cite{Esposito2016} &  &  & x &  & x & x &  &  &  &  & x & x &  &  & NR & 1.50 ppb & 25.00 \% & NR & 
         Medium & High & Medium & High  \\
         & RF & \cite{Borrego2018} & x & x & x &  & x & x & x &  &  & x & x & x &  &  & NR & 1.97 ppb & 17.00 \% & 0.89 & 
         Low & Low & Medium & Medium \\
         & FFNN & \cite{Borrego2018} & x & x & x &  & x & x & x &  &  & x & x & x &  &  & NR & 2.00 ppb & 25.00 \% & 0.81 & 
         Low & Low & Medium & Medium  \\
         & MLR & \cite{Cordero2018} &  &  & x &  & x & x &  &  &  &  & x &  &  &  & 2.92--3.88 ppb & 2.34--3.14 ppb & 12.3--35.4\% & 0.81--0.93 &
         N/A & Low & Medium & High \\
         & RF & \cite{Cordero2018} &  &  & x &  & x & x &  &  &  &  & x &  &  &  & 2.29--4.73 ppb & 1.86--3.35 ppb & 16.0--32.9\% & 0.85--0.95 & 
         N/A & Low & Medium & High  \\
         & SVM & \cite{Cordero2018} &  &  & x &  & x & x &  &  &  &  & x &  &  &  & 2.87--4.94 ppb & 2.07--4.36 ppb & 19.8--34.3\% & 0.79--0.95 &
         N/A & Low & Medium & High  \\
         & FFNN & \cite{Cordero2018} &  &  & x &  & x & x &  &  &  &  & x &  &  &  & 3.24--7.44 ppb & 2.76--6.22 ppb & 20.2--93.4\% & 0.62--0.88 & 
         N/A & Low & Medium & High \\
         & LR & \cite{Lin2015} &  &  &  &  & x &  &  &  &  &  &  &  &  &  & NR & NR & NR & 0.88 & 
         Low & Medium & N/A & High  \\
        \hline
        \multirow{2}{*}{NO\textsubscript{x}} & TDNN & \cite{Esposito2016} &  &  & x &  & x & x &  &  &  &  & x & x &  &  & NR & 1.37 ppb & 20.00 \% & NR & 
        Medium & High & High & High \\
         & FFNN & \cite{Esposito2016} &  &  & x &  & x & x &  &  &  &  & x & x &  &  & NR & 1.95 ppb & 29.00 \% & NR & 
         Medium & High & Low & High \\
        \hline
        \multirow{13}{*}{O\textsubscript{3}} & MLR & \cite{Maag2016} & x &  &  &  & x & x &  &  &  &  & x & x &  &  & 2.8 ppb & NR & NR & NR & 
        High & High & Medium & Medium  \\
         & FFNN & \cite{Spinelle2015} & x &  &  &  & x & x &  &  &  &  &  &  &  &  & NR & 1--4.5 ppb & NR & 0.915 & 
         Medium & High & Medium & Medium  \\
         & RF & \cite{Zimmerman2018} & x & x &  &  & x & x &  &  &  &  & x & x &  &  & NR & 0.7 ppb & NR & 0.99 & 
         Medium & Medium & High & Medium \\
         & MLR & \cite{Zimmerman2018} &  &  &  &  &  & x &  &  &  &  & x & x &  &  & NR & 5.1 ppb & NR & 0.81 & 
         Medium & Medium & Medium & High  \\
         & LR & \cite{Hasenfratz2012} &  &  &  &  &  & x &  &  &  &  &  &  &  &  & NR & 1.46 ppb & NR & NR & 
         Low & High & Medium & High \\
         & FFNN & \cite{Maag2018} &  &  &  &  &  & x &  & x &  &  & x &  &  &  & NR & 3.5 ppb & NR & 0.91 & 
         Medium & N/A & Medium & High \\
         & TDNN & \cite{Esposito2016} &  &  & x &  & x & x &  &  &  &  & x & x &  &  & NR & 7.45 ppb & 42.00 \% & NR & 
         Medium & High & Medium & High \\
         & RF & \cite{Borrego2018} & x & x & x &  & x & x & x &  &  & x & x & x &  &  & NR & 1.32 & 20.00 \% & 0.97 & 
         Low & Low & Medium & Medium  \\
         & FFNN & \cite{Esposito2016} &  &  & x &  & x & x &  &  &  &  & x & x &  &  & NR & 7.90 ppb & 70.00 \% & NR & 
         Medium & High & Medium & High  \\
         & LR & \cite{Saukh2015} &  &  &  &  &  & x &  &  &  &  &  &  &  &  & 12.9 ppb & NR & NR & NR & 
         Low & High & Low & High  \\
         & FFNN & \cite{Borrego2018} & x & x & x &  & x & x & x &  &  & x & x & x &  &  & NR & 2.60 ppb & 18.00 \% & 0.86 & 
         Low & Low & Medium & Medium \\
         & LR & \cite{Lin2015} &  &  &  &  &  & x &  &  &  &  &  &  &  &  & NR & NR & NR & 0.92 & 
         Low & Medium & N/A & High  \\
         & MLR & \cite{Maag2018} &  &  &  &  &  & x &  & x &  &  & x &  &  &  & NR & 10.7 ppb & NR & 0.16 &
         Medium & N/A & Medium & High  \\
        \hline
        \multirow{3}{*}{PM\textsubscript{2.5}} & FFNN + GP & \cite{Cheng2014} &  &  &  &  &  &  &  &  & x &  & x & x &  &  & 96.69 ug/m\textsuperscript{3} & NR & NR & NR & 
        Medium & High & High & High  \\
         & MLR & \cite{Liu2017a} &  &  &  &  &  &  &  &  & x &  &  &  &  & x & NR & NR & NR & 0.9959 & 
         Medium & High & N/A & High  \\
         & FFNN + GP & \cite{Gao2016} &  &  &  &  &  &  &  &  & x &  & x & x &  &  & NR & NR & 5.40 \% & NR & 
         N/A & High & N/A & High  \\
        \hline
        \multirow{2}{*}{SO\textsubscript{2}} & RF & \cite{Borrego2018} & x & x & x &  & x & x & x &  &  & x & x & x &  &  & NR & 0.09 ppb & 5.00 \% & 0.95 & 
        Low & Low & High & Medium  \\
         & FFNN & \cite{Borrego2018} & x & x & x &  & x & x & x &  &  & x & x & x &  &  & NR & 0.16 ppb & 10.00 \% & 0.86 & 
         Low & Low & Low & Medium \\
        \bottomrule
    \end{tabular}}
\color{black}
\end{sidewaystable*}

%% file: sections/roadmap.tex
\section{Discussion and Roadmap}
\label{sec:roadmap}

In this survey, we have critically compared common technologies and methods for machine-learning-based calibration of low-cost sensing units, including the sensing units themselves, machine learning algorithms used for constructing the calibration models, and evaluation measures for assessing the usefulness of the models.
We next reflect on the current state of the field, highlighting open issues that need addressing, and briefly presenting some directions for future research. As a basis for identifying open issues, we use the comparison of previous studies shown in Table~\ref{tab:appcomparison}.

\bparagraph{Combination of Sensors.} Considering that sensors have cross-sensitivities between pollutants, an important research problem is to find the best combinations of sensors that capture as many pollutants as possible without suffering in calibration performance. As also shown in Table~\ref{tab:appcomparison}, most studies currently focus either on calibrating gas sensors or particular matter counters, instead of attempting to calibrate both types of pollutants simultaneously. Capturing actionable pollution information requires accurate measurements for all pollutants included in relevant air quality indexes. 
A related topic is finding the best combinations of sensors so that they can complement each other, e.g., combining infrared and laser-based LSP sensors in the same sensing unit, to optimize energy consumption and accuracy while offsetting each others' disadvantages.
Indeed, most sensing units currently encapsulate a single sensor per pollutant instead of combining different types of sensors for the same pollutant.

\bparagraph{Life Cycle Management.} Massive-scale \acp{AQS} deployments are often built out of a heterogeneous base of sensors that are unattended and installed in hard-to-reach locations.
Routine tasks such as cleanup or software updates become hard to manage, which can lead to high maintenance costs.
Device life cycle management with minimal need for manual intervention is critical for continuous long-term operation of these deployments.
Achieving this with massive deployments is an open issue.
Another open problem related to life cycle management is detecting or predicting when a sensor has failed or is about to fail.
This can be potentially accomplished using ML techniques, but these techniques have not yet been investigated in the context of low-cost \acp{AQS}.

\bparagraph{Mobility Effects.} Mobility can significantly affect the accuracy of sensors.
When a sensor is in movement, the quantity of air that enters the sensor increases proportionally to its speed of travel, which in turn can increase the concentration of the pollutant detected by the sensor.
As we have already discussed, there are some ways to measure movement speed, so that it can be taken into account on the accuracy.

\bparagraph{Universal Models.} Most studies that attempt to calibrate multiple pollutants report mixed performance, with the best model differing for different pollutants.
The most evident example can be found in Cordero et al.
\cite{Cordero2018}, where no single model is the best performing for all gases and PM classes.
Developing models that perform well for multiple pollutants is currently an open issue.

\bparagraph{Deep Learning.} 
Little work has been done on applying deep learning for calibrating low-cost \acp{AQS}, with the only works we are aware of being those of Yu et al.~\cite{Yu2020_1, Yu2020_2}. While these have shown encouraging results, the lack of openly accessible standard test datasets makes it difficult to fully assess the benefits of deep learning in the context of air quality calibration. In particular, the many complexities of air quality measurements, including drift, autocorrelation, cross-sensitivities, and other sources of uncertainty, mean that the risk of overfitting is high. Indeed, deep learning typically requires a large number of measurements before the model converges, and in heterogeneous environments can easily overfit without this being easy to identify~\cite{pulkkinen2020understanding}. Another challenge with deep learning is to determine how to optimally represent and model the data. The works of Yu et al.~\cite{Yu2020_1, Yu2020_2} rely on so-called sequence-to-point modeling where the idea is to use a sliding window over input measurements to derive an aggregate point estimate for the window. These models can improve overall performance, but do so at the cost of temporal resolution and are best suited for deployments having frequent sampling rates (e.g., minute or even less). Understanding these kinds of trade-offs in modeling is another important research direction.

\bparagraph{Dataset Length.} The studies by Maag et al. and De Vito et al. \cite{Maag2016, Maag2018, vito2020}, to the best of our knowledge, are the only ones that use a test dataset longer than a year.
This is important for capturing seasonal phenomena and the effects of different weather patterns.
In the future, it would be important to see more studies that test the models for periods longer than a year, so that they are tested on different conditions.

\bparagraph{Concept Drift and Re-Calibration.} Concept drift has been widely reported as an issue for low-cost sensor technologies.
However, most studies have not been able to assess its effect due to using only short measurement periods.
Concept drift can be mitigated using periodic re-calibration or online training, where new training data is continuously used to improve model performance.
Developing methods for detecting drift and triggering these mitigation techniques are currently open issues.

\bparagraph{Other machine learning paradigms.} Machine learning paradigms, such as transfer learning or semi-supervised learning, could be readily utilized in sensor calibration. Transfer learning would be able to, e.g., speed up learning on new sensor hardware by utilizing commonalities with other domains, and semi-supervised learning could utilize the potentially large number of measurements for which calibrated ground truth does not exist.

\bparagraph{Use of Multiple Performance Measures.} Most studies use only one or two performance measures for evaluating the performance of calibration models.
Further work is needed to assess calibration accuracy using multiple, complementary measurements.
Also, the practical results of calibration have not been thoroughly assessed, with the study by Cheng et al.~\cite{Cheng2014} being the only one to consider calibration performance in practical applications.
Specifically, they consider how the results of the calibration model would affect the values of an air quality index.

\bparagraph{Virtual Sensor.} The dominant approach for air quality monitoring relies on low-cost sensors, a reference station, and a calibration model. 
Another possibility is to combine calibration with \textit{virtual sensors} where so-called proxy variables---variables that are correlated with the target variable---are used to construct a model that can be used to estimate the value of the target pollutant.
For example, we can measure BC concentration using low-cost PM sensors by building a model that can learn the relationship between PM and BC to estimate BC concentration from PM concentration~\cite{Zaidan2020}.
Using virtual sensors and proxy variables can reduce the required number of sensors, as fewer specialized sensors are needed, allowing for denser deployments or lower costs, or both.

\bparagraph{5G and Edge Computing.} High-resolution spatial-temporal air quality monitoring requires a dense deployment of low-cost sensors in an urban area, which brings many challenges for providing system support, such as ubiquitous high-speed connectivity and real-time analysis. 
Emerging 5G networks and edge computing deployments provide many of the necessary mechanisms, e.g., by offering faster data rates, energy-efficient networking, and having support for computing and data storage closer to the sensors. 
As an example of the potential benefits, Das et al.~\cite{das2020energy} report a 10-fold reduction in the energy costs of network transmissions when using 5G.
These improvements offer opportunities for new types of air quality monitoring solutions. For example, hyperspectral remote sensing has recently been used to estimate air quality~\cite{hou2020algorithm}. The improvements in networking technology offered by 5G and edge computing make it possible to deploy and use low-cost hyperspectral cameras for monitoring urban areas, e.g., by mounting the cameras on buildings or other infrastructure. Exploring novel modalities for air quality monitoring enabled by these improvements is another important research direction. 

%% file: sections/conclusion.tex
\section{Summary and Conclusion}
\label{sec:conclusion}

Low-cost air quality monitoring technology is emerging as a complementary technology to professional-grade air quality stations.
The high cost of professional-grade stations limits the granularity at which they can capture pollutant concentrations, whereas low-cost sensors can be deployed densely to increase the spatial granularity of collected information.
Unfortunately, the accuracy of low-cost sensors tends to be poor as the sensors are vulnerable to several sources of noise.
In this article, we have critically surveyed machine-learning-based calibration of low-cost air quality sensors, the main technique for improving the usefulness of measurements provided by low-cost air quality sensors.
Our focus has been on individual sensing units, each of which typically integrates several different sensors (e.g., environmental sensors, particulate matter sensors, and sensors for gaseous pollutants).
In this survey, we have covered the sensor technology itself, the processing pipeline required for calibration, the machine learning techniques that are used in calibration, and different ways to evaluate the performance of calibration models.
Based on our survey, we have highlighted open research issues in the field, with the inconsistency of studies, lack of sufficiently long datasets, and lack of models that perform well across several pollutants being among the most critical research problems.